*VIEW* ━━━━━━━━━ Communicated by Terrence Sejnowski

# Sensory Metrics of Neuromechanical Trust

**William Softky**
*wsoftky@stanford.edu*
*Bioengineering Deptartment, Stanford University, Menlo Park, CA 94025, U.S.A.*

**Criscillia Benford**
*criscillia.benford.gmail.com*
*Continuing Studies, Stanford University, Menlo Park, CA 94025, U.S.A.*

**Today digital sources supply a historically unprecedented component of human sensorimotor data, the consumption of which is correlated with poorly understood maladies such as Internet addiction disorder and Internet gaming disorder. Because both natural and digital sensorimotor data share common mathematical descriptions, one can quantify our informational sensorimotor needs using the signal processing metrics of entropy, noise, dimensionality, continuity, latency, and bandwidth. Such metrics describe in neutral terms the informational diet human brains require to self-calibrate, allowing individuals to maintain trusting relationships. With these metrics, we define the trust humans experience using the mathematical language of computational models, that is, as a primitive statistical algorithm processing finely grained sensorimotor data from neuromechanical interaction. This definition of *neuromechanical trust* implies that artificial sensorimotor inputs and interactions that attract low-level attention through frequent discontinuities and enhanced coherence will decalibrate a brain's representation of its world over the long term by violating the implicit statistical contract for which self-calibration evolved. Our hypersimplified mathematical understanding of human sensorimotor processing as multiscale, continuous-time vibratory interaction allows equally broad-brush descriptions of failure modes and solutions. For example, we model addiction in general as the result of homeostatic regulation gone awry in novel environments (sign reversal) and digital dependency as a sub-case in which the decalibration caused by digital sensorimotor data spurs yet more consumption of them. We predict that institutions can use these sensorimotor metrics to quantify media richness to improve employee well-being; that dyads and family-size groups will bond and heal best through low-latency, high-resolution multisensory interaction such as shared meals and reciprocated touch; and that individuals can improve sensory and sociosensory resolution through deliberate sensory reintegration practices. We conclude that we humans are the victims of our own success, our hands so skilled they**

**fill the world with captivating things, our eyes so innocent they follow eagerly.**

## 1 Introduction

Thanks to advances in communications infrastructure and interactive digital technology, digital stimuli constitute an ever-growing portion of human informational diets. Mobile digital devices, once niche products with limited versatility, are now ubiquitous workhorses and convenient sources of entertainment. They are an integral part of lived experience for many people around the world, providing unprecedented access to digital sensorimotor data and transforming our everyday understanding of what it means to connect with other human beings.

"The Internet" and "technology" are often blamed for changes in social norms and social expectations. Google is making us stupid. Our phones are turning us into narcissists. Our games are desensitizing us. Might hysterical nostalgia be fueling concern over the rising prominence of the Internet in our private and public lives? In some cases, yes. Nevertheless, a growing body of scientific research is dedicated to understanding how the Internet is changing individuals, groups, and culture more generally. In this body of research, excessive time online has been consistently correlated with psychosocial distress—namely, depression, loneliness, and anxiety. What accounts for this link? (For summaries of work in the field, see Carbonell, Guardiola, Fuster, Gil, & Tayana, 2016; Tokunaga & Rains, 2010, 2016; Cash, Rae, Steel, & Winkler, 2012; Block, 2008; Young, 1999.)

Internet addiction, defined as an impulse control disorder much like pathological gambling, was first proposed as a diagnosis for inclusion in the DSM in 1996 by the psychologist Kimberly S. Young (1996). The multidisciplinary field she created, though growing, still suffers from methodological limitations and lacks stable answers. Most important, What exactly is Internet "addiction"? A primary disorder of the reward system? A manifestation of an impulse control disorder? What mechanisms account for its co-occurrence with loneliness, depression, and social anxiety? Does Internet addiction evolve from psychosocial distress or vice versa? Are some individuals more prone than others? How best to treat it? Unfortunately, such methodological questions are compounded by dizzying multiplicities of variables—content, display, interactivity, incentives, users, ages, countries—together making consistent, stable data hard to find and prevalence rates hard to measure with consistency. We know that excessive time online is related to psychosocial distress, but we do not know why.

While researchers labor toward definitive answers, technology accelerates. Digital stimulus streams are becoming progressively easier to access and will likely continue to load 21st-century informational diets. This prospect raises two important research questions about the demands that digital interactivity and digital sensorimotor data place on the human nervous system: Do our informational diets matter? How? The value of these two questions lies precisely in their generality, because only a general framework can span present and future technologies and cross disciplinary boundaries.

The theoretical framework we introduce to explain how digital interactivity affects the human nervous system suggests that the structure of a human's informational diet matters. We use the term *informational diet* to denote the sensorimotor data a person consumes on a regular basis. We use the term *neuromechanical trust* to denote the statistical ur-computations underlying accurate world models. In our framework, sensorimotor data are differentiated primarily by data format, bias, latency, and bandwidth. Thus, we are not concerned with type of online activity (e.g., gaming, shopping, surfing) or content (e.g., violent, educational, political). Instead, we distinguish between the continuous, fractal-like coherence of natural sensorimotor data and interactivity versus the discontinuous, artificially enhanced coherence of digital data and interactivity. We wish to quantify both natural data and interactivity and digital data and interactivity in common terms relevant to the computations our brains perform.

This stripped-down version of "nature" and "the Internet" contributes to the generality of our theoretical framework. It is a framework inspired by three approaches well known elsewhere: (1) the theoretical physics practice of employing reference frames that invite difficult problems to surrender themselves to simple mathematical descriptions, (2) the hardware/software practice of tightly specifying data formats and interface protocols, and (3) the big data practice of projecting noisy, quantized, high-dimensional data onto smooth, low-dimensional curves. In our approach, the reference frame, the data format, and the smooth curves are all found in the structure of three-dimensional space-time. Our principles are theoretical: mathematics, physics, data science, signal processing, and control theory.

We use these assumptions to describe potential sensorimotor data in terms of informational quantity (entropy), coherence enrichment (statistical profile), and spatiotemporal qualities (latency) and to identify environmental factors relevant to the global phenomenon known as Internet addiction disorder. We also use these assumptions to model the human brain as a homeostatically self-calibrating circuit, informational diets as "calibration fields" containing various concentrations of noise and coherence, and circuit/field interactions as processes of information foraging. The homeostatic self-calibration model we present here, though it was originally conceived of independently, now revises a well-known information foraging model developed by Peter Pirolli and Stuart Card (1999), which itself revised earlier animal foraging models (Charnov, 1976).

Since Barlow (1961), neuroscience has maintained that the general purpose of the human perceptual system is to make accurate models of incoming signals (Hinton & Dayan, 1996). We understand these as *predictive* *perceptual models*. How exactly such models are produced remains an open question, as readers of this journal well know. What remains beyond question are two things: first, human perceptual models operate substantially like those of our primate ancestors, and second, such operation obeys mathematical constraints. Specifically, primate evolution established the statistical environment in which brains operate, and the mathematics of information flow and compression establishes how brains must use that information. These constraints are the terms of an implicit statistical contract between brains and the natural world.

We model human brains simply, as primate brains. We model homeostatic self-calibration even more simply, based on the foraging strategies of bacteria or birds. We believe a simple model has the best chance of capturing the general, low-level computational processes that underlie high-level perception. Our circuit model's simplicity means it should be regarded as a minimal concept map, not a biorealistic model. With it, we confirm and explain a well-known sign-reversal dynamic—a failure mode of homeostatic control systems known as homeostatic fragility, by which simple homeostatic systems become irreversibly entrapped by particular statistical environments (Ramsay & Woods, 2014). We dub this sign-reversal dynamic *leading-indicator dependency*.

The robustness of that particular entrapment cycle leads us to descriptively predict, on theoretical grounds, that dysfunctional Internet use is a maladaptive response to an informational diet high in coherence-enriched stimuli (abundant on the Internet), one leading to problematic cognitions that reinforce unconscious preferences for coherence-enriched stimuli. Left unchecked, this feedback loop will cause the human brain to lose trust in its senses, its companions, and itself. This loss of trust is progressively related to resolution loss in the brain's perceptual model. We also predict that the most efficient way to maintain and restore resolution, and hence brain function, will involve systematic continuous sensory exposure: to live (rather than recorded or synthesized), interactive (rather than broadcast), naturally coherent (rather than coherence-enriched), and multisensory (rather than monosensory) signal streams originating from three-dimensional space. Such streams are easily found in nature and human companionship.

We discuss this framework's possible role in guiding the digital technology industry to design products that are sensitive to human sensory needs. We also compare the framework's predictions to existing research and suggest future research directions. Finally, we discuss implications.

## 2  Our Theoretical Inspiration

We have reason to hope that a simple mathematical model can encompass human informational appetites because an even simpler model has already done much more by locating cognitive forces in the ancient foraging dynamics of bacteria. In a theoretical tour de force, Hills (2006) locates the principal axis of attention, and thus of human attentional disorders like attention deficit hyperactivity disorder, obsessive compulsive disorder, and autism, as a single, simple information-theoretic trade-off pitting focus against blur. Hills dissects the "area-restricted search" strategy by which microorganisms like bacteria loiter near food sources and uses computer-based genetic algorithms to illustrate the core one-dimensional strategy of these or any foraging creature, regardless of the space in which it forages: stay or go—that is, move slowly, if you find success ("stay"); move faster the less you find ("go"). This seems to be the simplest and perhaps the only way a creature could change its velocity to locate food in space: stay in place and sharpen spatial focus to select for lower search-space entropy versus go and blur spatial focus, like diffusion, to increase search entropy.

Theorists have acknowledged a related mathematical insight since Darwin: that the optimization performed by evolution and by genetic algorithms such as Hill's operate by alternating mutation/randomization/diffusion (i.e., by increasing entropy) with selection or selective amplification (i.e., reducing entropy). In alternating selection with diffusion, the theories of Darwin and Hills, in their mathematical essences shorn of molecular and historical detail, describe how informational organisms modulate entropy to optimize crucial variables. Darwin showed how animals evolve; Hills shows how we move.

These foraging dynamics have a special appeal to the algorithmically inclined: the sharpening information signal that bacteria pursue is authoritative and unbiased, being created by real gradients in the physical world. So although the forager's present input controls how it finds future input, that feedback loop passes through reality, so there is no danger of getting stuck in a self-reinforcing state. But that guarantee no longer holds for more complex creatures, for which information itself can be a resource.

Foraging for and storing information involve representation and anticipation (i.e., anticipating future resources in future places). As Hills (2006), explains: "Where once we would expect the animal to perseverate in the presence of the reward, we may now expect the animal to perseverate on the expectation (stimulated by external or internal cues) or 'idea' of the reward" (p. 16). Anticipation in particular involves trading away certainty to receive faster success, and thus more success, on average in the right statistical environment. But representation's fundamental trade can be a devil's bargain, because the forager itself must now replace the real sharpness of physical gradients with algorithmic sharpening, which requires selective amplification. Unfortunately, the stability of such a feedback loop is highly sensitive to input statistics. Modest changes in the input's accuracy—an accuracy that is no longer guaranteed—can wreak potentially catastrophic consequences on the behavior of such a simple circuit, as we show in the section on foraging theory.

## 3  Review of the Literature on Internet Addiction Disorder ____________

Individuals around the world are using the Internet in ways that have a
negative impact on their lives (Cheng & Li, 2014). The phenomenon goes
by many names in the literature, and there are at least 29 independently
developed measures of Internet addiction (Tokunaga & Rains, 2010). Some
studies conceptualize "the Internet" as the Internet in general, while oth-
ers focus on a specific Internet activity (e.g., gaming, shopping, interacting
with social media, interacting with pornography; Montag et al., 2015). Vari-
ations in conceptualizations of the medium itself and the behavior associ-
ated with it reflect disagreements regarding nomenclature and disparities
among key variables and variable valuation principles. These inconsisten-
cies limit the generalizability of existing research. Other limiting factors in-
clude sampling techniques and small sample sizes (Tokunaga & Rains, 2010,
2016; King, Delfabbro, Griffiths, & Gradisar, 2011).

Although the general concept of Internet dependence has gained credi-
bility within the social and behavioral sciences as meta-analyses and neu-
roimaging findings continue to show continuity among related constructs
(Yao & Zhong, 2014; Tokunaga & Rains, 2010; Young, 1999), neither "In-
ternet addiction disorder" (IAD) nor "problematic Internet use" (PIU) is a
diagnosis recognized by the American Psychiatric Association's *Diagnos-
tic and Statistical Manual* (DSM). "Internet gaming disorder" is listed in ap-
pendix of the DSM-5 as a condition warranting further research.

**3.1  Signs of Addiction/Problematic Use.**  Internet "addicts," as deter-
mined by validated diagnostic instruments such as Young's Internet Ad-
diction Test (IAT), typically present with problems regulating emotions,
impaired cognitive and interpretive skills (e.g., decreased sensitivity to
nonverbal messages), psychosocial distress (e.g., loneliness, depression,
anxiety), and reward sensitivity (Carbonell et al., 2016; Swingle, 2015;
Bipeta, Yerramilli, Karredla, & Gopinath, 2015; Young, 1999). Risk factors af-
fecting the speed and likelihood of developing clinically significant levels of
Internet dependence include boredom, shyness, marital discontent, work-
related stress, financial insecurity, general insecurity, and a limited offline
social life (Young & Rodgers, 1998; Tokunaga & Rains, 2010). Comorbidities
include ADHD, chemical and behavioral addictions, obsessive compulsive
disorder, hostility, and depression (Young & Rodgers, 1998; Bipeta et al.,
2015). The causal direction of these relationships remains to be definitively
determined, perhaps because internet dependence is a cyclical rather than a
linear process (Yao & Zhong, 2014; Caplan, 2003). Yet despite controversies,
there are consistent findings, which we summarize.

**3.2  Neuroimaging.**  Neuroimaging studies show molecular, morpho-
logical, and circuit-level similarities in the brains of individuals suffer-
ing from Internet dependence, substance dependence, and pathological gaming, suggesting overlapping mechanisms at work in all three phenomena (Hong, Zalesky, et al., 2013; Hong, Kim et al., 2013; Dong, Devito, Du, & Cui, 2012; Lin et al., 2012; Yuan, Qin, Liu, & Tian, 2011; Yuan, Qin, Wang, et al., 2011; Ko et al., 2009). Prolonged recreational Internet use has been correlated with both task-related and resting-state abnormalities, including abnormalities in gray matter volume and white matter fractional anisotropy (FA) (Weng et al., 2013). These abnormalities correspond to observed behavioral impairments in executive control, mental flexibility, emotional regulation, response inhibition, goal-directed behavior, motivation and reward processing, higher-order cognitive functions, and memory encoding and retrieval. (For systematic review of neuroimaging studies to 2012 see Kuss & Griffiths, 2012, and Brand, Young, & Laier, 2014.)

**3.3 Impulse Control Disorder Tradition.** Dysfunctional internet use parallels substance dependency in several ways (Li & Chung, 2006; Zhou et al., 2011). However, the more influential models of dysfunctional Internet use see it as an impulse control disorder much like pathological gambling; cognitive-behavioral therapy is recommended for treatment (Davis, 2001; Young, 1996). Generally defined, impulsiveness names a predisposition to react quickly to stimuli (external or internal) without regard for possible negative outcomes (Moeller, Barratt, Dougherty, Schmitz, & Swann, 2001). An individual suffering from pathological impulse control would tend to act quickly and with little forethought in pursuit of "short-term gains that accompany significant long-term costs" (Tokunaga & Rains, 2016). The Internet is a particularly attractive stimulus in this context because, like slot machines, its interactive nature supports the kind of unpredictable variable reward structures used in classical instrumental conditioning processes (Young & Nabuco de Abreu, 2010; Rossi & Yin, 2012).

Impulse control disorders involve psychological as well as physiological dependence on the stimulus (Le Moal & Koob, 2007). Examples from the DSM category "impulse-control disorders not elsewhere classified" include compulsive gambling, kleptomania, trichotillomania, and pyromania, among others (Holden, 2001). Factors generally understood to be part of the impulse control model of problematic Internet use include viewing the Internet as a source of instant gratification, euphoric sensations and other rewards stemming from Internet use, tolerance, preoccupation with the Internet, withdrawal, long-term negative outcomes traceable to use, and continued use of the Internet to escape aversive mood states (Tokunaga & Rains, 2016; Cash et al., 2012; Young, 1999; Lesieur & Rosenthal, 1991; Chou & Hsiao, 2000).

**3.4 Uses and Gratification Tradition.** Research in the uses and gratification tradition conceives of the Internet as an accessible and attractive communicative context because it offers escape from feelings of loneliness or depression (Yao & Zhong, 2014; Young & Rodgers, 1998; Davis, 2001).

Users can find other users any time of day or night, making the Internet particularly useful for avoiding feelings of loneliness. It offers anonymity, psychological distance, and structure—as well as increased control. Online social success leads to positive alteration of mood (Caplan, 2003).

Online socializing, however, does not seem a sufficient substitute for offline socializing (Chou & Hsiao, 2000), and users who prefer the Internet as a communicative context over face-to-face interaction may be at higher risk of developing Internet addiction (Caplan, 2003; Davis, 2001). Internet use expectancies (e.g., anticipating escape from an aversive mood) and poor coping skills have also been found to be reliable predictors of addiction (Brand et al., 2014). Yao and Zhong (2014) confirmed findings by Chou and Hsio (2000) demonstrating that while feelings of loneliness and other symptoms of Internet addiction were reduced by offline socializing, an increase of online socializing neutralized the effect. Moreover, online socializing by itself did not reduce feelings of loneliness. Their findings suggest a "worrisome vicious cycle between loneliness and Internet addiction" (Yao & Zhong, 2014).

**3.5 Prevalence Rates.**  A 2014 meta-analysis of 31 nations in seven world regions obtained an overall prevalence estimate of 6% with moderate heterogeneity. Following Young (2009), the authors defined what they termed generalized Internet addiction (GIA) as "an impulse control problem characterized by an inability to inhibit Internet use that exerts an adverse impact on major life domains (e.g., interpersonal relations, physical health; Cheng & Li, 2014). The authors conclude that, taken together, their findings provide "tentative support" for the hypothesis that GIA prevalence is related to quality of offline life, as reflected in indicators of subjective assessment (life satisfaction) and objective quality (environmental conditions such as traffic congestion, overall pollution, physical safety, and purchasing power). The regions with high prevalence rates were more unpleasant, with lower life satisfaction rates and "greater overall pollution (primarily air pollution), greater traffic commute time consumption, and lower national income." The analysis obtained a 10.9% prevalence rate in the Middle East (the highest in the world) and a 2.6% prevalence rate in Northern and Western Europe (the lowest). The prevalence in North America was estimated to be 8% (Cheng & Li, 2014). These data refute the common assumption that prevalence rates are caused by Internet accessibility and align with data in both the "uses and gratifications" and "impulse control" traditions that see online life as an escape from offline difficulties.

As noted earlier, worldwide prevalence rate estimations vary. Cash et al.'s (2012) summary of research in the field to 2012 reports that prevalence rate estimates for the United States and Europe range tenfold, from 1.5% up to 18.5%. These uncertainties underscore the importance of meta-analytic work.

## 4  Framework Overview

It is helpful to review basic tenets of information processing for readers from other disciplines. The concept most central to this framework is the structure of continuous three-dimensional space-time. This construct supplies the reference frame and data format that constitute our representational paradigm and the principles by which we convert sensorimotor stimuli into numerical data. We also turn to the principles of mathematics, physics, data science, signal processing, and control theory to further refine our characterization of available sensorimotor data, describing potential stimulus streams in terms of data format (as noted above), informational quantity (entropy), coherence enrichment (statistical profile), and spatio temporal qualities (latency). We use this same characterization of sensorimotor data to arrange potential environments on a continuum ordered from least to most nourishing in Figures 11 and 13 in the appendix.

We summarize the key terms below. An asterisk denotes further explanation and quantification in the appendix.

**4.1  Principles versus Neurons.*** This framework's foundational principles do not refer to circuit elements caricatured from observations (discrete neurons and connections) but rather to mathematical laws: laws of mechanical resonance, active control, continuous representation, and information transfer. In principle, such laws might together circumscribe a "zero-parameter" understanding of the human body/mind system based primarily on math, and only secondarily on experimental evidence. Such an understanding would be both powerful and neutral.

**4.2  Attention.** Inspired by Hills, we understand attention as the strategy used by an active sensorimotor system to allocate computational resources to the space it represents.

**4.3  Continuous versus Categorical Representations.*** Perception and learning in traditional neuroscience usually refer to categorical distinctions, such as object types, decisions, or behaviors. But a brain's twin tasks of millisecond muscular coordination and the continuous representation of reality require vastly more and better-structured information than categories could ever provide. In our view, while categories can transport sensory information from the natural world, they are bad at representing it because the natural world is continuous.

**4.4  The Unconscious.*** Implicit and explicit information processing are different (Frith & Frith, 2008, 2003). Implicit ("unconscious") information processing denotes a prereflexive mode that is faster and less controllable than explicit information processing ("conscious"), which is rule based, more controlled, and reflexive. These two forms of processing are best conceptualized as a continuum (Fiske & Neuberg, 1990). Computational limits on consciousness indicate that the vast majority of sensory and sensorimotor processing is implicit and inaccessible to conscious recollection.

**4.5 Neuromechanical Trust and Empathy.** The word *trust* has meaning in computer science, or rather two meanings. First, a computer can trust its input data without authentication (binary signal-trust), as in the skipped validation stages of Figure 2. Or a computer can trust its model of an external system (analog model-trust, as in equation 4.1 and Figure 4). Both of these concepts are behind our conception of neuromechanical trust.

No one needs to tell you that your hand is connected to your elbow or your feet are on the ground. You can *feel* it. That is neuromechanical trust—the trust your nervous system has in its model of mechanical configuration, which makes the world seem stable and predictable, a trust statistically assembled from countless micro-interactions. Neuromechanical trust in micro-saccades and coordinated vibrations of the eyeballs stabilizes vision so perfectly it is unconscious. Neuromechanical trust in active mechanical control of the cochlea (of the sort responsible for tinnitus) allows us to localize sound sources to a few fingers-width in azimuth and elevation, better than most other animals. Neuromechanical trust in whole-body micro-tremors makes those vibrations imperceptible to us until one holds laser pointers or binoculars.

The signal-processing metric of latency jitter is a good marker for neuromechanical trust, and hence for trust in general. And the metric of neuromechanical trust is a good marker for social trust, which arises from the same inputs, outputs, and circuits as does neuromechanical trust, and so ought to be described by the same concepts.

Neuromechanical trust is the most information-dense and reliable form of trust possible, atop which derivative forms of trust can then be built using the same statistical algorithms and often the same sensors. As a data-intensive sensorimotor experience, neuromechanical trust requires physical proximity so that the signals from all sensory modalities are maximally informative and consistent. Like an authentication value in computer science, neuromechanical trust must be reestablished with each connection; it cannot be simply inherited or copied. Because establishing neuromechanical trust requires continuous, authentic sensory signals, it cannot arise from representations, categories, or digitized inputs. It can only be built on proximity experience.

Neuromechanical trust in other people constitutes empathy. We predict that one's capacity for empathy, "the ability to comprehend another's feelings and to re-experience them oneself" (Salovey & Mayer, 1990), is correlated with one's capacity for neuromechanical trust in other beings. We treat all forms of trust, including interpersonal trust, as lower-bandwidth versions of the statistical trust one has in one's balance or one's senses. The advantage of neuromechanical trust as a clinical measure rests with its relative ease and neutrality of measurement.

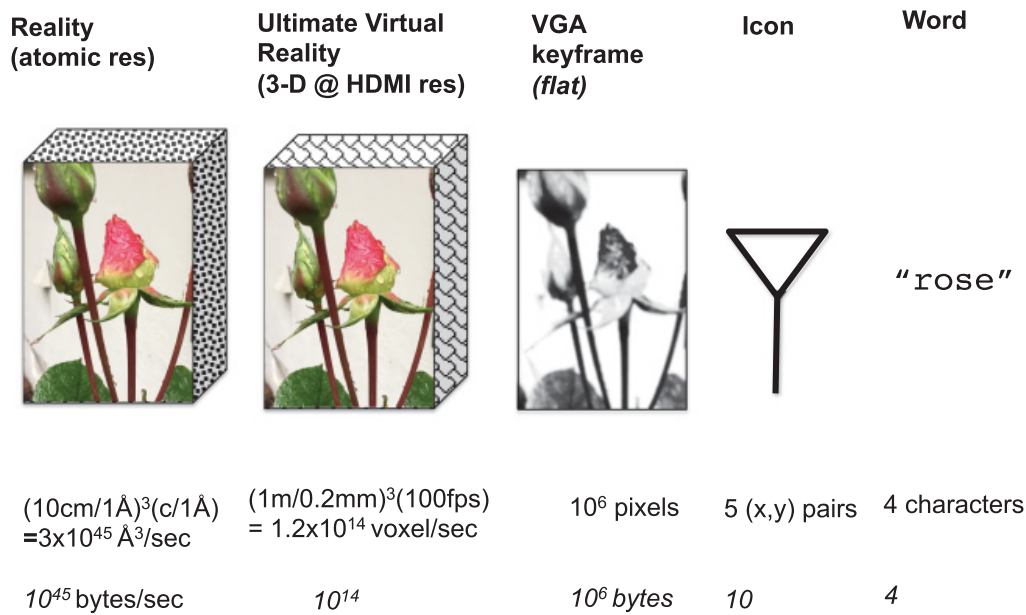

Figure 1: Compressed versions of reality. At atomic resolution, an actual rose, fixed in time, could be said to contain $10^{45}$ bytes/sec of information. But depending on the resolution and format, a compressed representation of the same rose could occupy from $10^{14}$ bytes/sec down to 4 bytes.

Empathy can thus be considered a physical resonance practice and the only possible cure for loneliness.

### 4.6 Compressed Representations.

A picture can be worth a thousand words or, equivalently, a few kilobytes—that is, 5000 characters at 1 byte each or, equivalently, a medium-sized, slightly pixelated image. Such metrics and such principles underlie all our results, because both sensory and digital processing involve the same compression trade-offs between input data flow, output resolution, and statistical assumptions linking them.

Discrete digital information can sometimes be compressed without loss, but continuous signals cannot. For example, consider a rose in a garden (see Figure 1). The original rose contains effectively infinite information, or at least as much information as the arrangement of its atoms. The same rose modeled volumetrically, with volume-elements (i.e., voxels) the size of typical pixels, would use a tiny fraction of the atomic scale value, and coarser representations would use even less, all the way down to the few bytes needed for a word or icon, whose savings come from dispensing with the original detail.

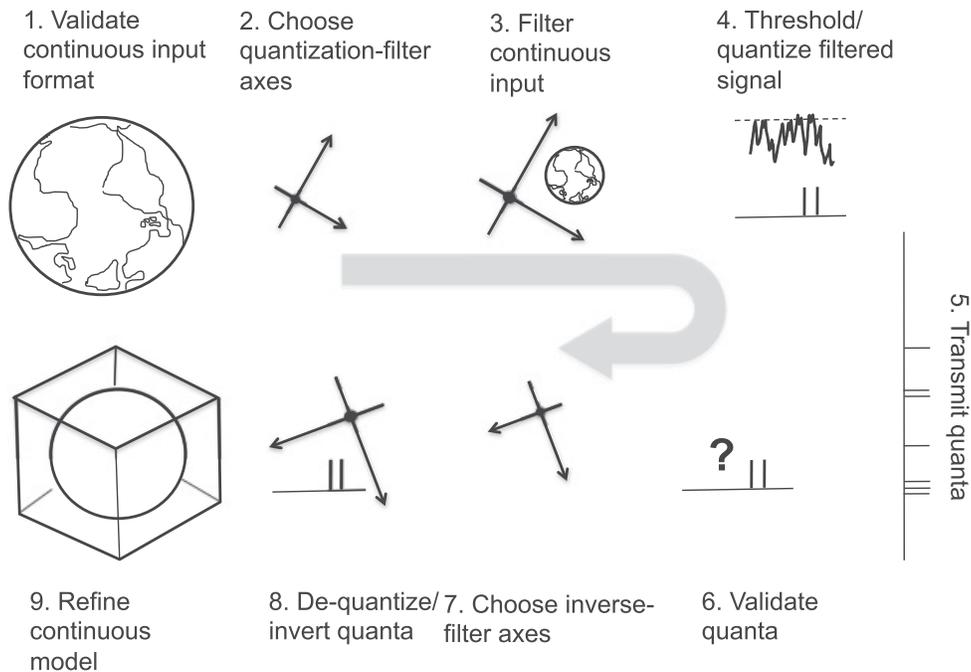

Figure 2: Steps for encoding information from continuous fields. Each numbered step describes a part of a system for converting continuous information into quantized form and back again, compressing and then uncompressing it. In systems optimized for efficiency, signal trust is assumed, so authentication steps 1, 2, 6, and 7 are skipped, forming a potential security vulnerability.

**4.7 Bayesian Priors and Assumptions.** One cannot analyze data or decompress signals without making statistical assumptions, and the strength of those a priori assumptions about signal patterns directly affects the resolution of decompressed object representation. Assumptions must be made about the shapes and patterns to be found, based on statistical redundancies (i.e., the structure) known to be present in natural images and objects (Olshausen, 2016). By narrowing the space of allowed patterns, such priors proportionally improve resolution if enforced during decompression.

For example, driving the tens of millions of pixels on a high-definition screen would nominally require 1 gigabyte per second of data if individual pixels were statistically independent. But by making assumptions about smooth, continuous contours and contiguous surfaces, the decompression technology in HDTV devices creates beautifully realistic images using a relative trickle (less than 1%) of that raw information flow.

**4.8 Compression and Decompression Occur in Stages.\*** Those who frequently use digital devices are already familiar with the central concepts underlying compression, in particular, encoding, transmission, and decoding of streamed images, as shown in Figure 2.

**4.9 Bandwidth.*** The quality of a representation depends on its information content, or rate in bytes per second, so sensory bandwidth limits the quality of sensory representations. The lowest-bandwidth senses are smell and taste, because they change so slowly; we leave them out of our analysis. In descending bandwidth order, our top three channels are touch, vision, and sound. Their collective bandwidth ranges from about 10 kBit up to 10 Mbit, depending on whether the neural spikes are averaged together in a rate code or treated separately in a pulse code.

**4.10 Noise.*** Variability in nature is noisy and random, and can be averaged into better signals. Variability in the digital world, in contrast, was sculpted just for us, in that the structure of its variability, such as flickering pixels, can either take advantage of or block our signal-averaging abilities. The process of encoding or digitizing automatically designates some signals as noise, but that choice is arbitrary. As Carver Mead once said, "One man's noise is another man's information" (Softky, 2014).

**4.11 Diffusion and Sharpening.** Diffusion, when applied to the particles in a distribution, blurs or coarsens the distribution and thus lowers the information it can convey. It needs no extra energy, only thermal jostling. A good mathematical approximation involves convolving the distribution with a local-averaging function such as a gaussian bell curve. Paradoxically, at the particle level, diffusion is the most entropy-increasing process possible, but its net result of smoothing lowers the entropy of patterns that can be represented in the medium.

Sharpening is the opposite of diffusion but is rarer because (like other forms of amplification) it needs extra energy. Mathematically, sharpening enhances the local derivative of a signal instead of its local average.

**4.12 Signal Formats and Metrics.*** Sensorimotor data and interactions are perhaps the most important theoretical constructs in this framework. In information technology, valid signals must have the proper format to be trustworthy. Brains expect sensory spikes from continuous natural sources in proximity and expect that signals from all sensory modalities will arrive in synchrony, unfractured. Because the nervous system compresses and decompresses signals from those sources during sensorimotor activity, we can apply traditional signal-processing metrics to quantify both natural sensorimotor and digital signals, even those lacking the proper format. In Figure 3 and (Table 2 in the appendix) we illustrate eight such metrics for sensorimotor input as a template for the more complex sensorimotor interactions that actually occur.

**4.13 Media and Interface Distortions.*** An interface lies between the organic sensorimotor system and the calibration field. Originally that interface was air. We consider air as neutral because it is consistent, transmitting signals with maximum fidelity along each of the eight metrics of Figure 3. Relative to it, digital communications are nonneutral in many ways, in particular by decreased bandwidth, lower temporal fidelity, and more interruptions. (Figure 13 and Table 3 in the appendix quantify many of these distortions for digital technologies.)

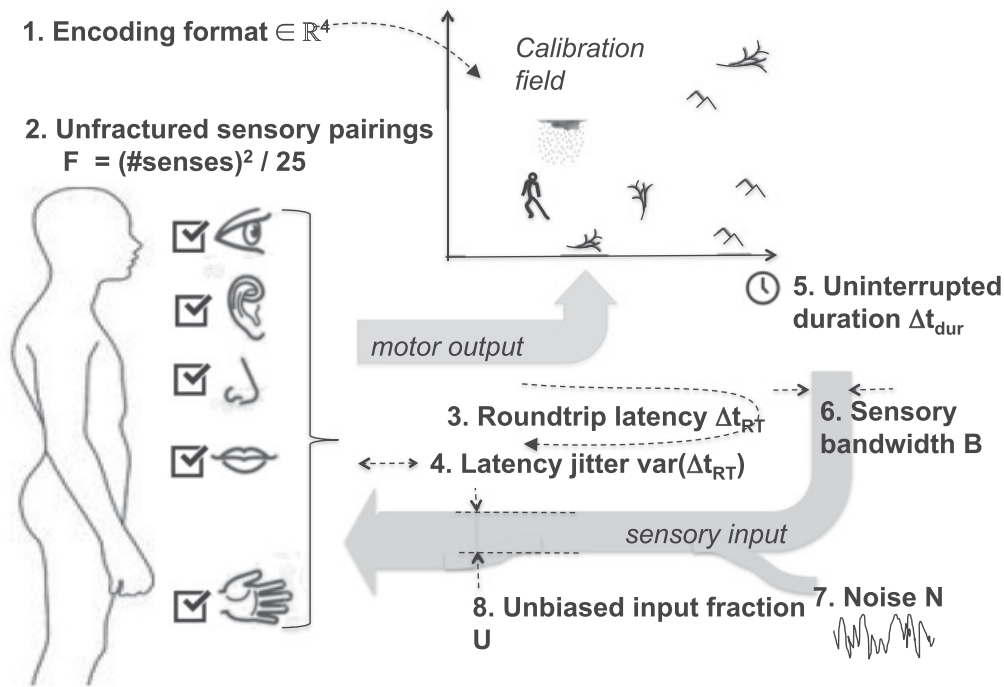

Figure 3: Interface metrics described over a continuous embedding space. Each metric is a scalar correlated with information flow. (1) A human calibration field must be topologically consistent with our native 3D data format to obtain a nonzero value for this metric. (2) After encoding format, the most important requirement of calibration signals is that they involve as many senses as possible. An easy estimate of cross-sensory interaction is given by the square of the number of senses involved, so a measure of 1.0 describes full sensory fusion and 0.04 a monosensory stream. The other metrics shown are well known in signal processing: (3) round-trip latency from motor signal to response; (4) moment-to-moment jitter in that latency; (5) the maximum duration of calibration episodes; (6) the bandwidth of the sensory return signal; (7) the noise added to the signal; and (8) the fraction of the input that is not selected or biased by the interface. Not each metric will matter in every circumstance, but all matter sometime. Digital media fare far worse on all metrics than unimpeded access to the natural calibration field.

**4.14  Natural Statistics.*** The neuroscientific concept of natural statistics distinguishes the fractal complexity of natural scenes from the simplified structure of man-made cages and buildings. We explore this distinction in depth in section 5.

**4.15 Prediction Engine Models.** Neuroscience has long described the brain's primary task as modeling reality. The model created can be conceived of as a particular form of informational structure: a real-time, unsupervised prediction engine that correlates prior knowledge of the spatiotemporal structure of the world with advance-correlated signals coming from it, signals we term *leading indicators*. By "prediction engine," we do not mean a specific set of neural pathways. Rather, the term denotes the computational principles by which the algorithmic processing of neural impulses predicts the immediate high-dimensional sensory future (as if predicting pixels), and from the prediction-reality mismatch generates a high-dimensional error signal (Softky, 1996; Seth, 2015).

Prediction as a process has two requirements: good priors and good data. Bayesian priors are the statistical assumptions underlying any data analysis. For a brain, they could be strong assumptions such as three-dimensional continuous motion (3-D priors) (Softky, 2014), moderate assumptions such as Marr's (2010) 2.5-D sketchpad, weak assumptions such as the hierarchical segmentation of deep learning (LeCun, Benglo, & Hinton, 2015), or the assumption of continuous sequences (Cui, Ahmad, & Hawkins, 2016). Regardless, the accuracy of a brain's predictions will still in each case also be proportional to the quality and quantity of the sensorimotor data available to it.

**4.16 Calibration as a Process.\*** Instruments are calibrated by engineers and scientists to maintain resolution, noise immunity, and robustness. But strictly speaking, calibration applies less to the physical instrument than to a model of its input/output relations. So in the broadest sense, the goal of calibration is to produce a trustworthy model, operating over as wide a range and with as small an error as possible. In this sense, a trustworthy model would roughly maximize model trust, which we can represent relationally as follows:

$$Model\ trust = \frac{model\ range}{model\ prediction\ error}. \tag{4.1}$$

Our understanding of calibration in general is summarized in Figure 4.

**4.17 Self-Calibration.** We model brains as high-resolution instruments requiring continuous tuning and calibration via learning and plasticity. But as with many learning models (and unlike most instruments), a brain's calibration must be unsupervised, so it must forage through space for calibration signals while foraging for survival, using similar strategies. This calibration must occur at multiple timescales simultaneously—for example, stabilizing millisecond-scale balance at the same time as year-scale social relationships. While our approach describes sensorimotor learning in terms of signal processing and information theory, substantially the same concepts are involved in learning "sensorimotor contingencies" (O'Regan & Noe, 2001). Our models are described in detail in the section 6.

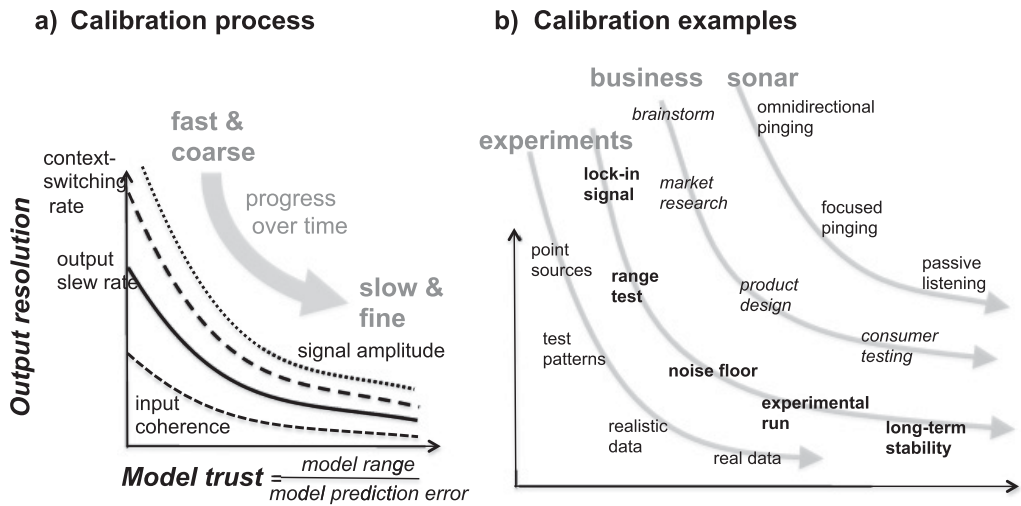

Figure 4:  Calibration. (A) Process. Because calibration tends to improve model resolution and the resolution of the model and of the instrument's outputs should be commensurate, one expects various output measures (e.g., amplitude) to decrease over time. (B) Examples of well-known calibration processes.

**4.18  Palatability.\***  We assume informational organisms will seek to recalibrate by seeking clean, unambiguous signals in natural environments. That is, they will find such signals attractive and palatable in some informational sense like "interestingness" (Schmidhuber, 2009) or "curiosity" (Little & Sommer, 2013). We map this appetite for clean sensory data to the concept of attractiveness or "palatability" (Lutter & Nestler, 2009; Na, Morris, Johnson, Beltz, & Johnson, 2006).

**4.19  Sign-Reversing Feedback Loops.\***  When organisms are operating outside the statistical environments in which they evolved, their expected homeostatic balance can be upset, and stable negative feedback can be inverted into positive feedback in a situation called sign reversal.

Maladaptive states like addiction can result when an instinct to correct a problem makes the problem worse, thereby turning an initially stable negative-feedback loop into a self-reinforcing positive-feedback one, a catastrophic process that a survey in *Psychological Review* dubs "homeostatic fragility" (Ramsay & Woods, 2014). The review describes how various biological feedback loops can change their signs in environments not anticipated by evolution—loops such as energy balance (Strubbe & van Dijk, 2002), salt concentration (Na et al., 2006), or opioid analgesia (Slot, Pauwels, & Colpaert, 2002). Ramsay and Woods (2014) clearly state the central theoretical issue: "Health problems such as drug addiction and obesity (as well as other disorders of modern society) may result from homeostatic fragility exposed by evolutionarily unanticipated perturbations." As their first example, the authors cite "an effector response that eventually overcorrects an initial perturbation (i.e. causes a sign-reversal) despite the continued presence of the disturbance." We fear that digital addictions manifest this simple principle.

**4.20 Evolutionary Pressures.** The human nervous system has adapted to best survive in a natural environment, presumably in the direction of optimized fitness. Artificial signals, both physical and cultural, have evolved far more quickly, rewarded by fitness metrics such as palatability, cost, productivity, easy of transmission, durability, and so on. Both forms of evolution constitute optimization, and both produce optima that by definition have low entropy. But the optimum of human DNA enhances our ability to process a naturally noisy environment, while the optimum of artificial signals creates a low-entropy environment unfamiliar to our nervous systems.

## 5 Metrics of Trust for Natural and Artificial Inputs

**5.1 The Structure of Calibration Fields.** We begin by discussing the space from which the data originate—their "embedding space"—which we assume is three-dimensional continuous space over continuous time (i.e., $\mathbb{R}^4$). The primary structural constraints on the calibration field are topological (e.g., continuous and connected) and dimensional (e.g., 3D). Human brains, like all other brains, operate three-dimensional host bodies in three-dimensional environments. For most of our time on this earth, our brains have regularly consumed sensorimotor data originating in continuous space-time, a four-dimensional "real world" described by real numbers. Brains evolved long ago to process only data encoded in this format.

Screen-based displays do not transmit signals with this format, however fine their resolution. Mathematically, a million-pixel screen does not show a three-dimensional scene, but rather a two-dimensional projection of a three-dimensional sub-manifold of a million-dimensional space of pixels. Algorithms exist for discovering low-dimensional curved submanifolds embedded in high-dimensional spaces (Roweis & Saul, 2000; Tenenbaum, de Silva, & Langford, 2000), but the need for such algorithms and the difficulties of their implementation confirm that those two spaces are not equivalent.

We next derive the continuous statistical properties of calibration fields from the physical patterns most often found in nature: temporal patterns, spatiotemporal patterns, natural entropy, and social patterns.

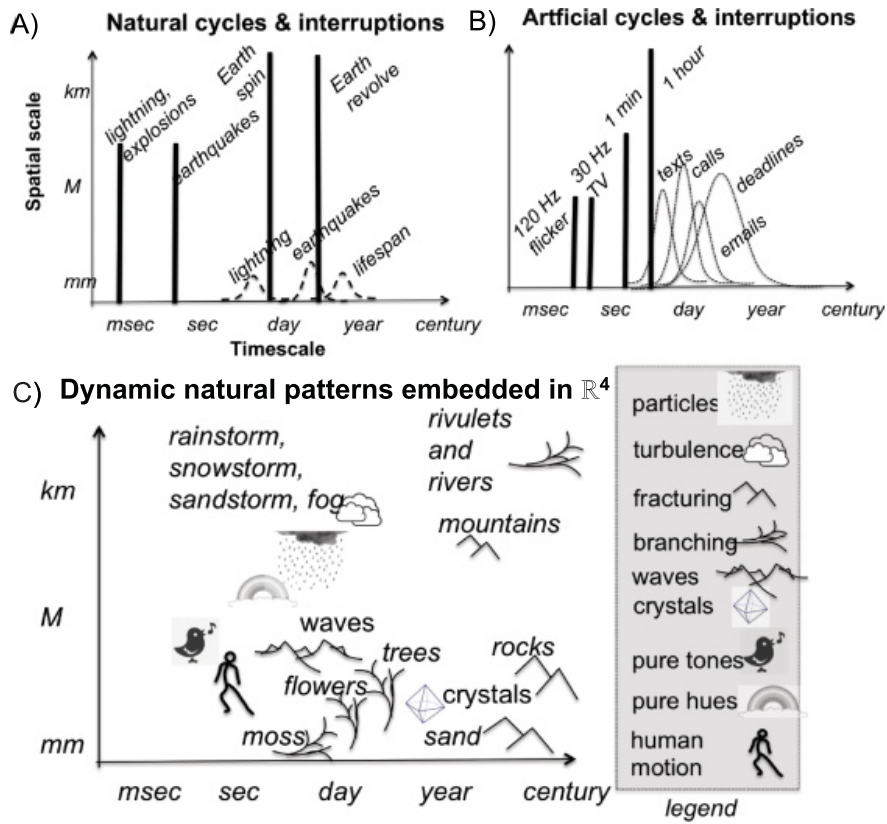

Figure 5: Spatiotemporal scales and patterns of the calibration field. Natural calibration fields are almost entirely continuous in space and time, spanning millimeters to kilometers and milliseconds to years. (A) Natural cycles and interruptions include regular oscillations such as days and years (delta function frequencies), along with rare, unpredictable interruptions such as lighting (bell curve timescales). (B) Artificial cycles and interruptions include faster divisions of time and far more unpredictable interruptions. (C) Dynamic natural patterns embedded in 4D space-time. Several types of natural fractal patterns, such as branching and fracturing, span orders of magnitude in both space and time (time-changing fractals could include a fern unfurling or a rivulet spreading). The "pattern" of a tree species corresponds to its statistical profile of root density, twig size, and branching angles rather than to the specific detail of any particular tree.

**5.2 Temporal Patterns.** Natural temporal patterns are continuous in time and often predictable, from microsecond pressure fluctuations like sound to millennia-long climate changes and even longer astronomical cycles. Among those trends and cycles occur occasional sudden events like eruptions and lightning, but even those arise from continuous equations of motion; no natural events are truly discontinuous, utterly unpredictable interruptions (see Figure 5A). While modern life also contains regular cycles such as 60 Hz alternating current and 60 minute hours, it also contains discontinuous interruptions such as phone calls and messages (see Figure 5B), which cannot be predicted from locally available signals. The extreme hyperpredictability of narrow-band vibrations can decalibrate broadband sensory circuits, and the absolute unpredictability of remote interruptions can trigger discontinuity-detecting and alerting circuits.

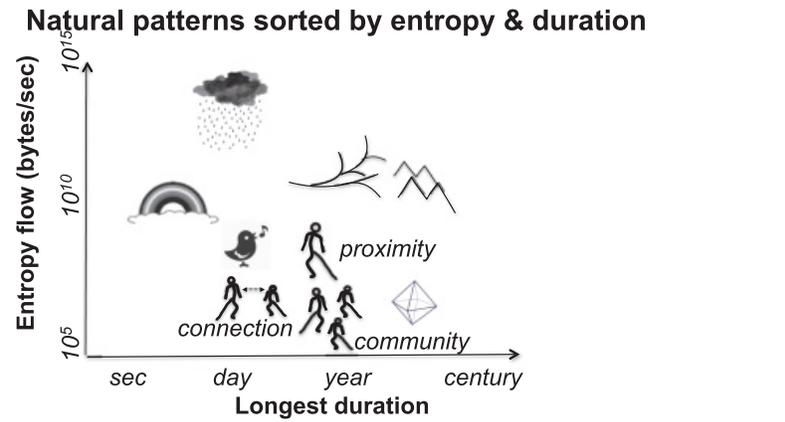

Figure 6: Natural patterns sorted by entropy and duration. Spatial patterns in the forest or savanna are ambiguous, complex, detailed, and diverse (i.e., of high entropy). In that environment, simple low-entropy patterns are rare and often fleeting, but those provide the cleanest signals.

**5.3 Spatiotemporal Patterns.** Spatiotemporal sources in the natural world are mostly fractal, at least to first-order approximation, so the patterns emanating from them will be too. Figure 5C caricatures the structure of spatial and spatiotemporal patterns in the natural world, sorted by space- and timescale. All of these patterns are continuous in space. Most have the highly irregular statistics of particles (e.g., fog droplets) or fractally corrugated surfaces (e.g., moss, trees, waves, mountains). Artificial spatiotemporal patterns appear in such a wide variety of man-made media, ranging from asphalt sidewalks to smartphone screens, that their description is more diverse (see the appendix). In general, their display format is low-dimensional (screens or speakers), their structure is ultra-high-dimensional and quantized (pixels and frames), and the principles of their construction are biased toward influencing the human nervous system.

**5.4 Natural Entropy Patterns.** Figure 6 sorts the same natural sources shown above—rivers, clouds, gravel and such—according to entropy, plotting the most cluttered patterns (e.g., raindrops) at the top and the least cluttered patterns at the bottom. Our self-calibration analysis says that humans should find attractive low-entropy natural patterns such as crystals, with pure colors, truly flat faces, uniform composition, and smooth surfaces.

In general, artificial patterns (both physical and digital) have far lower entropy than natural ones because they arrive via the iterated low-entropy-amplification filter of optimization, which over time produces some equilibrium distribution of artificial patterns. One can guess at that distribution's structure. A useful analog is the distribution of active modes in a laser cavity (e.g., TEM-00). Of course cultural norms are not photons, and a media echo chamber is not a Fabry-Perot mirror cavity. But both amplify information selectively, both are resource constrained, and both experience mode competition. Lasers produce coherent, monochromatic light, the lowest-entropy electromagnetic patterns possible. Material communication produces materially productive physical patterns, also of low entropy compared to the natural variability of human communication. Evidence of low-entropy modern environments is as close as the nearest flat wall or floor (Barrett, 2014). (See the appendix for a detailed comparison of the entropy of natural versus artificial patterns.)

**5.5 Social Patterns.** Social patterns are the significant subset of natural sources and sounds that emanate not from rocks and trees but from fellow creatures. Though incredibly intricate, that manifold would contain human-specific spatiotemporal structures (gait, posture, gesture) useful for calibration and communication at all timescales from milliseconds to years (Vogeley & Bente, 2010). While human motion and social interaction seem incredibly complex on their own, in fact they have relatively low entropy compared to turbulence or jumbled sticks and stones and thus provide signals specially tuned to our own nervous systems.

In general, natural communications are vibratory, high bandwidth, and temporally precise, while artificial ones are quantized, low bandwidth, and desynchronized. Because social trust arises from the same inputs and circuits as does neuromechanical trust, what degrades one will degrade the other.

## 6  Foraging Models of Self-Calibration

We assume that creatures evolved to find the information they need. Research conducted by Bromberg-Martin and Hikosaka (2009) shows that "the same dopamine neurons that signal primitive rewards like food and water also signal the cognitive reward of advance information." They conclude that "current theories of reward-seeking must be revised to include information-seeking." We build on this generalized concept of informational reward to understand how a creature can not only forage for useful information, but can keep itself in a state of perpetual self-calibration by means of homeostatic mechanisms (which nonetheless may be brittle because of homeostatic fragility). We use off-the-shelf techniques from other disciplines, in particular, foraging theory (from economics and behavioral biology) and control loop theory. The key assumption is that any circuit will have evolved to match a particular statistical environmental profile and thus will implicitly expect its environment to obey that statistical contract.

**6.1 Foraging Theory Background.** The term *information foraging* was coined by Pirolli and Card (1999) when they adapted an evolutionary biology model of optimal food foraging to predict the behavior of humans who forage for information. Developed by Eric Charnov (1976), the original model of animal foraging behavior represents the patchy resource distributions characteristic of nature; the mathematics representing this behavior is known as the marginal value theorem (MVT). Pirolli and Card (1999) used MVT to identify those factors influencing the timing of movement from one resource patch to another.

We adapt their model for several reasons. Pirolli and Card (1999) conceptualize information in the colloquial sense: information contained in words and documents, that is, information that is quantized, static, and consciously accessible. This common conceptualization of information does not permit comparison between natural and artificial information sources. Moreover, as we have shown, the magnitude of such information is tiny compared to the Shannon information transmitted by natural sources.

Charnov's canonical model represents possible principles governing the search for resources in a physical environment using "birds" that hop between discrete "bushes" full of berries (see Figure 7a). A bird decides to leave one bush for another when the marginal benefit (including switching costs) exceeds the benefit of staying on the present one. Charnov's basic hypothesis was that **"**when feasible, natural information systems evolve toward stable states that maximize gains of valuable information per unit cost." This approach sees information as a resource distributed in space, to be sought by spatial foraging strategies. Such maximization leads to near-optimal hopping, but intermittent hopping cannot implement the continuous control we seek. Nor does the berry bush model of resources adequately approximate the reduced costs involved in online information foraging. Finally, maximizing anything is different from tracking metabolic needs.

Gazzaley and Rosen (2016) have already applied foraging theory to Internet addiction. They did so in order to "explain our seemingly constant need to switch our focused attention in response to either an external alert or internal trigger"—even when we know that such switching creates "resumption lags" that have a negative impact on productivity. In their adaptation, different online portals (e.g., mobile phones, tablets, laptops) represent different berry bushes. They use their model to knit together empirical work on multitasking, showing that the behavior is motivated by both internal and external factors.

Crucially, they differentiate between the 3-space in which our informational appetites evolved and the $N$-space of the Internet. The difference lies in the mathematical way distance is measured (i.e., in the structure of metric spaces). In the real world (i.e., 3-space), interesting things take time to

**a) Foraging for resources**

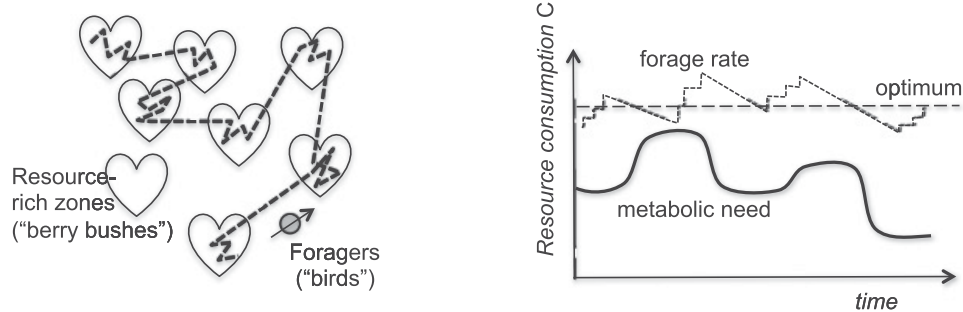

**b) Navigating a resource gradient**

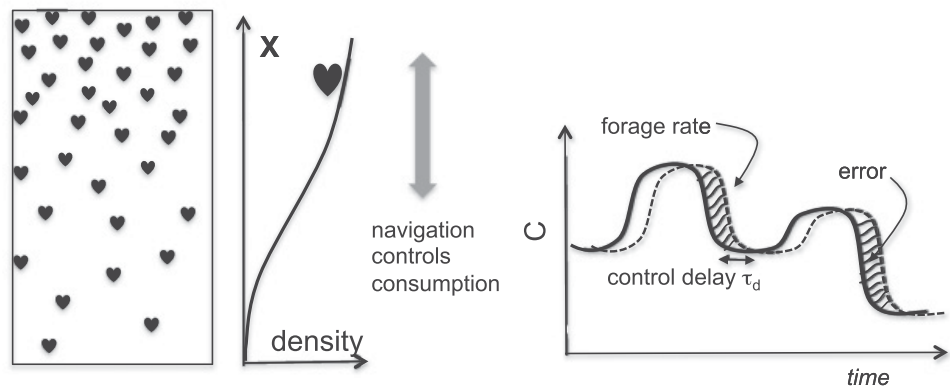

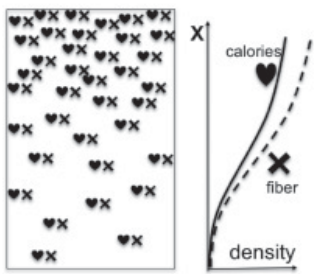

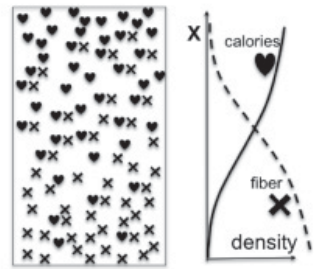

Figure 7: Models of resource foraging and control. (a) The canonical foraging model is of birds eating and depleting berries from bushes. This model maximizes consumption and does not track metabolic needs. (b) Our alternative postulates a continuous gradient of resource density through which a forager can navigate to meet its needs via homeostatic control. (c) To accommodate metabolic needs for multiple resource types (e.g., "calories" and "fiber"), both resource types may be co-joined with matching density gradients. (d) An alternative resource gradient is of enrichment, in which the relative fraction of the two resources changes over space. These distribution statistics prove problematic, being different in one crucial way from those the circuit evolved to process.

find, because as you move a distance $R$, the region in which you can search increases at most with $R^2$ (corresponding to surface area of an ordinary sphere). In contrast, on the Internet, instant gratification is but a click away in any one of thousands of directions (dimension $N = 1000$), not just the three directions we are used to. At each of those thousand new locations, another thousand branch off, ad nauseum. One can perform the same surface area calculation as before, merely by replacing the three-dimensional sphere ($N = 3$) with a hypersphere ($N = 1000$). The surface area of a hypersphere increases with $R^{N-1}$. This is far faster than the $R^2$ our nervous systems expect. The importance of this difference cannot be overstated. Gazzaley and Rosen (2016) conclude, as we do, that the excess of immediate confirmation signals available online hijacks our native mechanisms for attention management.

Yet because we model both behavior and nature as continuous in all respects, our approach differs from Gazzaley and Rosen's (2016) in modeling foraging as feedback control (rather than discrete decisions) and informational resource distribution as gradients (rather than patches).

To show how a general class of simple biorealistic circuits, whose structure and hardwired parameters originally "evolved" in one statistical environment, can become unstable when immersed in a different statistical environment, we offer two related sets of models: one set describing environments and other describing the control circuits that navigate in them. We model the environment in three progressively more complex ways. First is as a foraging field containing a single quickly measurable resource whose density various over space. We dub this resource "calories" to capture the quickness of the measurement (the sense of taste registers immediately). Our next two foraging fields contain two related resources, "calories" along with a new resource we dub "fiber," whose defining characteristic is that it can only be measured slowly, in the sense that roughage can be felt in the gut after some time has passed. In our second foraging field, these two resources have a fixed concentration ratio, whose joint density varies over space (see Figure 7c). Our third and final foraging field also contains those same two resources, but now with a fixed overall density, whose concentration ratio varies over space (see Figure 7d).

In each of those environments, our continuous foraging circuit uses closed-loop control to implement simple foraging. This is a toy model, as applicable to metabolic drives as to informational ones. The mechanism we highlight is simple: the circuits' negative-feedback loops, adapted in one statistical environment to be self-correcting and stable, convert in a different statistical environment into positive-feedback loops that are self-amplifying and unstable. In effect, we highlight statistically definable combinations of circuitry and environment prone to become addicted to ever quicker, coarse responses. Because this effect shares self-reinforcing dynamics with addiction and is driven by quickly measured signals, we call it "leading indicator dependency."

## 6.2 Environment Models: Density Gradient and Concentration Gradient

*6.2.1 Models.* In our model, the berry-eating bird resides in a world where resources are distributed along a smooth local gradient measured in "calories." There are no discrete berry bushes, only a continuous berry field, changing in berry density smoothly along its length, from sparse at one end up to bountiful at the other (see Figure 7b).

We now modify this baseline model to add a single new resource: "fiber." In our simplified world, we assume that fiber and calories exist with some concentration ratio **K**. In one scenario (see Figure 7c), **K** is fixed across space, so that the gradients of calories and fiber are always the same, even as density varies. We term this the "natural scenario." In the second scenario (see Figure 7d), **K** varies over the field while density is fixed. We call this the "enriched scenario."

*6.2.2 Principles.* Our berry-eating bird is stupid; it does not rest if it feels sated but instead merely moves to eat in a lower-density region. In the language of control theory, our bird's effectors controlling feeding speed share one fixed rate, so that only the spatial location effector may vary. This means that the forager has only one degree of freedom: it controls its calorie consumption and navigates in space by following the local calorie gradient. Thus, regardless of the various resources or their distribution profile, it always navigates using the local cue of calorie concentration.

## 6.3 One-Parameter Foraging model (Density Gradient).

A simple circuit can illustrate the principles of homeostatic control, measuring the single parameter calories (as measured during ingestion) to regulate the creature's feeding rate.

*6.3.1 Model and Principles.* Our circuit (see Figure 8a) is a classical proportional-integral-differential (PID) controller, as introduced by Schenck (1987, equation 5) to explain biological homeostatis. This PID controller makes a consumption variable, **C**, match a reference need, **N**, in the simplest way:

$$\frac{d}{dt}C = K_P \ (N - C) + K_D \frac{d}{dt} \ (N - C) + K_I \int \ (N - C)dt. \tag{6.1}$$

For our continuous forager, the reference value $N_{cal}$ represents the creature's current metabolic need (in calories), and its consumption $C_{cal}$ represents its attempt to meet that need, delayed by the (early) timescale $\tau_{cal}$ at which caloric intake can be measured. Unlike in many control-theoretic treatments, here we do not prove stability of the control loop, but assume it, in order to understand possible instabilities later.

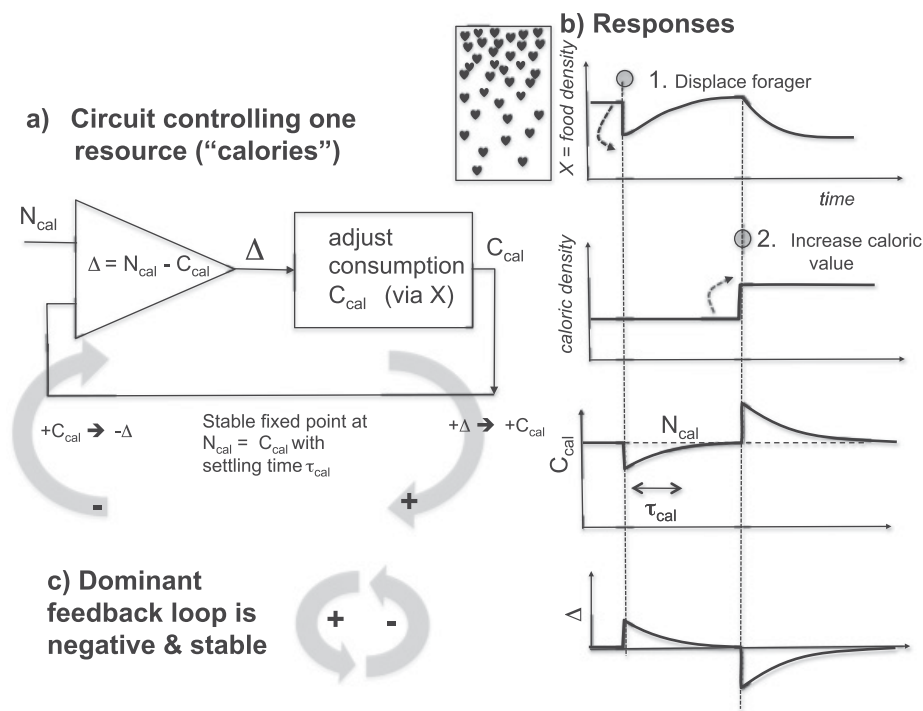

Figure 8: A simple control loop controls consumption via navigation. (a) The block diagram of a PID controller to make consumption **C** match metabolic need **N** (measured in calories). (b) The circuit finds new equilbria in response to perturbations: (1) displacing the forager creates a calorie mismatch, which drives it to return to its starting point, while (2) changing the overall calorie density forces the forager to forage in a different location. (c) This feedback loop is stable because it combines amplification with subtraction.

Typically, the coefficients ($K_P$, $K_D$, and $K_I$) are tuned to optimize the accuracy and lag of control, but here we need only a general description of when this process succeeds versus fails. The key ingredient for success is the self-correcting negative feedback loop: as $C$ rises, the three difference terms ($N - C$) reduce below zero, limiting further increase of $C$ (see Figure 8c). According to *Psychological Review*'s most recent summary of controversies in homeostatic regulation, this negative sign is crucial: "Indeed, a fundamental premise of any regulatory system is that deviations (or anticipated deviations) of regulated variables are detected and trigger effector responses that counter (or mitigate) the deviation" (Ramsay & Woods, 2014). Few references even discuss what might happen if the feedback became positive.

*6.3.2 Observations.* The loop can be seen as stable by its response to perturbations (see Figure 8b). If the forager is moved to a new location (the first event in Figure 8b), its newly modified appetites will return it to the original location. If one modifies the environment to decrease the berries' caloric density (the second event), then soon, within $\tau_{cal}$ , the forager will move—this time permanently—to a new location restoring the former calorie consumption—the expected behavior for such a control loop.

**6.4 Two-Parameter Foraging Model (Density Gradient).** The single-variable control loop above meets a single need, (in this case, "calories"), but real creatures have multiple needs and face trade-offs in measuring and acting on them all simultaneously. In an environment with a common gradient of two resources (calories and fiber), a forager chasing calories is automatically chasing fiber as well, so that the forager can successfully translate the need for fiber into an appetite for calories.

*6.4.1 Model.* Our circuit already measures calorie ingestion quickly, with timescale $\tau_{cal}$ as it navigates the local calorie gradient. Now, in addition, we allow for the forager's actual need $N_{cal}$ to be measured separately and at a slower timescale $\tau_{met}$ by the metabolic system as a ground truth. The forager is still using the same strategy as before, but now its inputs and processing have changed.

To accommodate this slightly more complex problem space, we modify our forager toy model (see Figure 9) by adding four features. The first change is distinguishing two distinct measurement timescales: the early but approximate one that measures calories in real time versus the delayed but authoritative timescale of confirmed metabolic needs after the fact. The second modification is to add a parallel measurement of the need for fiber, $N_{fiber}$, also measured with slower $\tau_{met}$. The third is to create a free-standing system that estimates the calories-to-fiber ratio $K_{est}$ of the food supply in order to calculate how many calories it must consume ($N'_{fiber}$) to meet the fiber needs. The final modification is to select the maximum of the twin needs—for calories and for fiber—as the single input to the existing calorie-only circuit. This last modification forges the perfect compromise, using each input according to its particular specialty and compensating for its weakness.

So we have two interacting circuits: outermost, a slow and ultimately unimpeachable verification of the forager's overall need for food; innermost, a much faster circuit that can find and consume the food it needs in real time using its fast-acting sense of taste. This circuit is optimal in several senses: (1) it uses the simplest control circuit for its core task (move and consume); (2) its reference signals ($N_{cal}$ and $N_{fiber}$) are accurate models of the forager's ultimate needs; (3) it uses its own metabolic-best-fit estimate ($K_{est}$) to meet those needs by controlling only one parameter; (4) it converges the single-parameter loop as quickly as possible by using a low-latency but approximate signal ($N_{cal}$ at $\tau_{cal}$); and (5) it couples the two circuits and timescales minimally to avoid complex interactions. These circuit properties could be considered optimal because they use reasonable assumptions about the natural environment—that calorie consumption tracks fiber consumption, or that berry composition remains uniform—to combine multiple, imperfect data sources into a collectively reliable stream. We cannot say this toy model is natural, but it is naturalistic insofar as nature tends to converge on simple and optimal solutions.

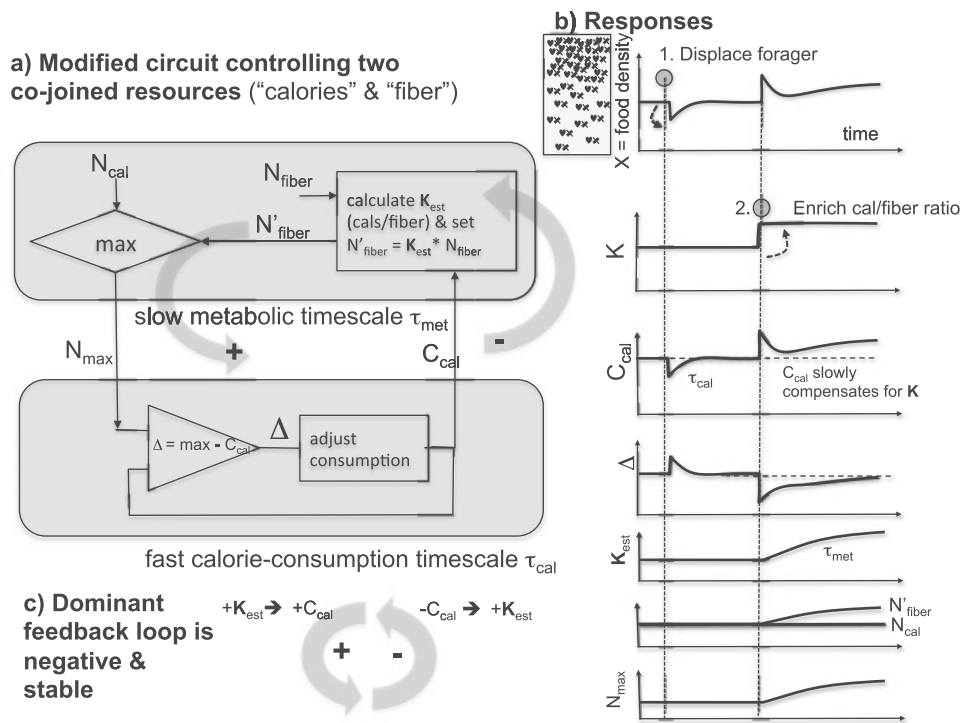

Figure 9: Modified loop controls two co-joined resources whose ratio is fixed. (a) The same circuit as before, now modified to forage for the greater of two separate metabolic needs. (b) This circuit exhibits stable equilibrium in response to changes in either parameter (location or density). (c) The stable response to perturbations in density occurs, as before, because subtractive interactions produce negative closed-loop feedback.

*6.4.2 Principles.* It is hard to overestimate the importance of simplicity to the performance of such a circuit. Foraging requires response speed, response speed requires throughput, and throughput requires simple computations on readily available signals with maximal assumptions and minimal validation. In information processing, any slowdown costs too much. Here, both simplicity and speed come from using leading indicators, a business term for signals advance-correlated with metrics of interest.

Leading indicators are such a useful processing trick that nature seems to have wired them to be learned from experience, as explained by the homeostasis survey: "Perhaps the most compelling examples of anticipatory regulatory strategies involve the use of initially arbitrary cues that, by virtue of prior associative experience, become conditioned to elicit corrective responses" (Ramsay & Woods, 2014). In fact, Hills's (2006) theory of goal-directed search finds support for learning leading indicators not only from foraging theory, but from detailed agreement with empirical work in algorithmic modeling, molecular genetics, and behavioral ecology: "The evidence strongly supports the evolution of goal-directed cognition out of mechanisms initially in control of spatial foraging but, through increasing cortical connections, eventually used to forage for information."

*6.4.3 Observations.* We can show this circuit is stable, as before, by adding perturbations. In the traces of Figure 9d, the first event displaces the forager, and as before, the forager quickly returns to its point of homeostatic equilibrium. The second event changes the nutritional composition of the berry field (more calories, less fiber), and as before, the forager moves to a new location that meets its needs using the new food supply. This second move has two components, one fast and one slow, moving in opposite directions. When the calorie density first rises, consumption spikes above demand ($\mathbf{C}_{cal}$ trace), so that the forager's first response is to eat less. Unfortunately, reducing food intake reduces fiber intake even further (less fiber per calorie in the new food multiplied by less food eaten). Fortunately, the metabolic loop slowly discovers the deficit as it calculates $\mathbf{K}_{est}$, thereby boosting $\mathbf{N'}_{fiber}$, and the nonlinear max() function lets this term dominate the consumption circuit to ingest more food instead of less. Such stable self-correction across environmental parameters is how this sort of a circuit ought to behave.

This success comes from using $\mathbf{C}_{cal}$ as a leading indicator, which converges faster than waiting for more accurate but slower signals like $\mathbf{N'}_{fiber}$. This trick is possible only because, in this environment, the consumptions of fast calories and slow fiber are correlated, so a single circuit measuring a single resource can guarantee consumption floors for both resource types at once.

The two-parameter toy circuit above seems nearly ideal for our two-parameter toy environment because it uses minimal modifications to a simple, well-characterized control circuit to merge fast, coarse information ($\mathbf{C}_{cal}$) with slow, precise information ($\mathbf{N'}_{fiber}$ and $\mathbf{N}_{cal}$). But that assumption depends on the statistical profile of the toy environment, in particular, that its enrichment has no spatial gradient. What happens if that constraint is lifted, so that the calorie gradient is different from the fiber gradient?

## 6.5 Two-Parameter Foraging Model (Enrichment Gradient)

*6.5.1 Model and Principles.* The circuit model remains the same as above, but in the new, enriched environment, on one side of the field are calorie-enriched berries and, on the other, fiber-enriched berries, so that berry density is fixed while berry concentration varies (see Figure 10). That means the bird, when it navigates toward calorie-enriched berries, thus navigates away from fiber-enriched berries. Eating more calories now reduces fiber consumption instead of increasing it. The circuit has stayed the same, but the environment has flipped the sign of the control loop.

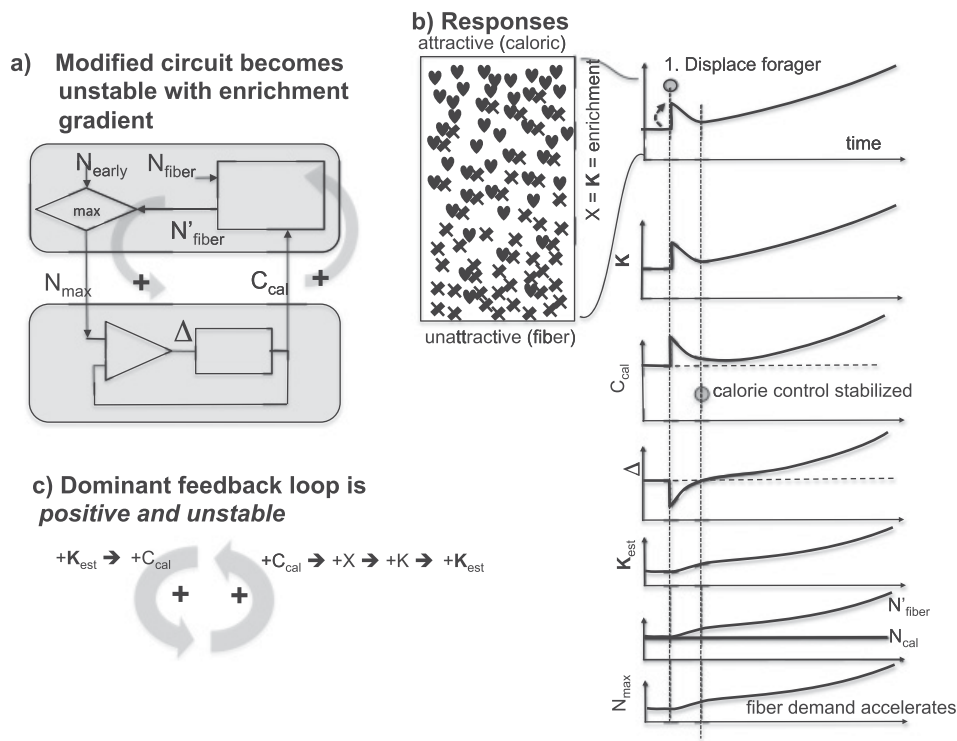

Figure 10: Enrichment gradient creates catastrophic instability. (a) The same modified circuit as before, but now foraging where high-calorie and high-fiber resources are separated in space, so that foraging for calories reduces fiber intake. (b) Responses to displacing the forager (1) toward higher-calorie food at first compensate for the immediate overconsumption of calories, but because fiber intake declines, this response exacerbates the problem. (c) This instability is not merely a failure to maintain control, but a critical conversion from a stable negative-feedback control loop into a highly unstable positive-feedback one: increased consumption $C_{cal}$ increases $X$, and hence $K$, and hence $K_{est}$, in turn increasing $C_{cal}$ further.

*6.5.2 Observations.* This model produces sign reversal as a result of principled design choices. The circuit's attempt to increase fiber consumption by increasing calorie consumption proves to be doomed in this environment. Consuming more calories now means consuming less fiber. As shown in Figure 10, the enrichment gradient makes previously stable circuit properties become unstable: the more calories the forager finds (increasing $K$), the hungrier it becomes, ad infinitum. The rising curves at left reinforce each other, so the circuit has no stable equilibrium point.

*6.5.3 Leading Indicators May Inform Homeostatic Sign Reversal and Salience Attribution.* The sign reversal we find differs in several key ways from the best-known model of sign-reversal mechanisms in biological systems (Slot et al., 2002). That simple numerical model has no spatial navigation and controls a single parameter, as in our Figure 8. It allows an arbitrary constant $w$ (not necessarily 1.0) in calculating the control signal (our $\Delta$) and finds sign-reversal effects for a seemingly unrealistic value of $w = 4$. Our model has no free parameters, it compresses two quasi-independent signals into one, and it explicitly navigates across palatability gradients.

But the similarities are revealing: both models involve homeostatic loops gone awry in evolutionarily novel environments, both subtract long-term averages from current values, and both involve conflicts between fast- versus slow-resolving measurements of crucial parameters. Such an astonishing overlap of mechanisms, inferred from disparate systems yet united in their addictive outcomes, cries out for more thorough mathematical understanding: Which timescales, gains, and gradients matter most in these feedback circuits? What are the circuits' phase-plot contours encompassing regions of homeostasis, runaway amplification, and saddle points? Need there be a gradient of palatability in the environment to induce these effects, or are other deviations from evolutionary statistics sufficient? Most critical, under what circumstances can circuits hard-wired by nature to pursue fast-resolving signals ever avoid falling into feedback traps? With such an understanding, we may approach a theoretical consensus on the mathematical drivers of addiction in general.

Simple homeostatic control provides an impoverished model of instrumentally conditioned behavior. In control, motor output is directly driven by a scalar input, while behavior can be decoupled from input by aversive signals or supervisory regulation. Nonetheless, our model can approximately be mapped to instrumental conditioning. As Rossi and Yin (2012) described, instrumental conditioning is a training process for generating and assessing the behavior of animals, often mice. They distinguish between goal-directed and habitual behaviors. Goal-directed behavior requires less training than habitual behavior and is more sensitive to "value and contingency manipulations" than habitual behavior. For example, instrumental conditioning can be used to train mice to associate lever pressing with the "reward" of food pellets. If sustained for long enough, random interval training, which introduces uncertainty in the process, can make the association between the stimulus (the lever) and the reward (food) "habitual," and therefore difficult to extinguish. Moreover, a mouse that has a "habit" of pressing the lever will continue to do so even after it has gone through a reward devaluation procedure. Morrison, Bamkole, and Nicola (2015) found that sign tracking behavior in mice—that is, the propensity to engage with the conditioned stimulus (in this case the lever) rather than the site of the reward—was "enhanced, rather than diminished, following reward devaluation."

Our model accounts for this behavior in terms of timescale by treating the conditioned stimulus as a variety of leading indicator. From the perspective of the simple control circuit, the leading indicator is not just faster, but also assumed to be authoritative. It replaces the reward signal itself in the name of efficiency. However, without having both separate variables, reversing the initial association is computationally impossible. Thus, if the circuit "knows" a reward is coming thanks to the leading indicator but that reward fails to appear, the circuit has no reference to identify the source of failure.

**6.6 Leading-Indicator Dependency in Enriched Environments.** These same fast/slow trade-offs affect self-calibration. Calibration is vastly more complex than single-variable foraging, involving multiple sensory modalities (sight, sound, touch), multidimensional inputs, complex environments, sequenced behaviors, and competing informational needs across space, time, signal intensity, coherence, and so on. Fortunately, multidimensional systems can still use homeostatis. As Koob and LeMoal (1997) explain, "The concept of homeostatis contends that an organism maintains equilibrium in all of its systems, including the brain reward system, that is, the organism uses physiological and cognitive or behavioral capabilities to function within the appropriate limits of physiology with the help of its own resources."

More important than the true complexity of the homeostatic system for understanding self-calibration as information foraging is the fact that while calories simply exist, information exist only through change in time, so a self-calibrating system would not only forage for signals it finds palatable (fast- versus slow-resolving sources) but would also for the sake of bandwidth have an innate preference for fast-resolving, high-bandwidth signals and the capacity to trigger them with its outputs. Such a temporally self-modulating circuit is beyond our simple theory here.

*6.6.1 Model and Principles.* To model self-calibration as information foraging, we appeal to Figure 4, which caricatured the process of calibration as a navigation toward subtler signals in search of greater model resolution. The motivating principle of calibration—and the statistical contract protecting homeostatic self-calibration from homeostatic fragility—is that fast-resolving, clean signals improve model resolution and thus constitute a leading indicator worth pursuing. Figure 11 represents a thought experiment designed to illustrate how self-calibrating systems could suffer from leading indicator addiction in that pursuit. The figure represents a toy environment with only two types of informational resources: fast-resolving signals (e.g., clean ones) and slow-resolving signals (i.e., noisy). The fast signals are like sugar, easy and quick to measure but not exactly what the forager needs, while noisy signals take time to resolve. In a natural environment, we suppose that these two high-amplitude signals occur simultaneously, so that looking for clean signals also provides noisy ones. In that equivalence "clean = calories" and "noisy = fiber," exactly as shown in Figure 11. In a natural, neutral environment of rare fast-resolving signals, this foraging strategy would succeed. But in an enriched environment, any deficit in model resolution will press the forager toward even stronger, quicker, less unambiguous signal sources, even if the problem results from too many of them already.

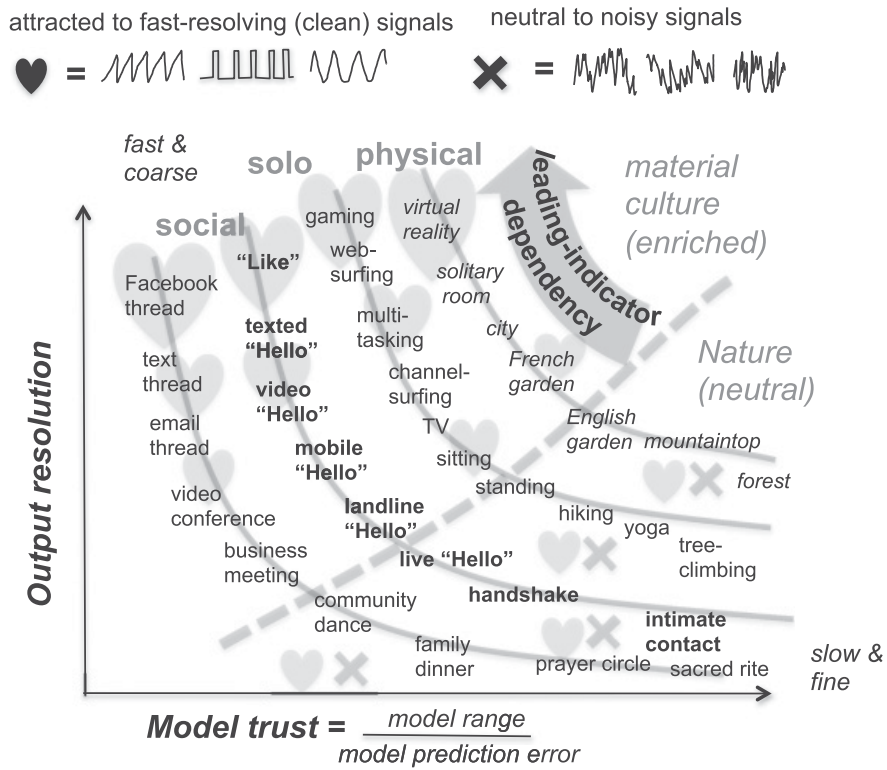

Figure 11: Modern environments are enriched with fast-resolving data, and hence likely to induce leading indicator dependency. In this graph, with the same form as Figure 4, varieties of slow-resolving sensory data and their social and environmental contexts are found on the lower right. On the upper left are varieties of quick-resolving (enriched) data. These data are often available in technologically influenced contexts; they are attractive to conscious drives but provide little long-term benefit in terms of model resolution. Indulging in them leads to an unconscious sense of decalibration, leading the homeostatic self-calibration circuit to seek yet quicker, cleaner input, fruitlessly course-correcting as in the runaway calorie/fiber dynamic.

## 7  A Note on Physical Media and Alternate Realities

Why didn't novels and television produce the leading indicator dependency we describe here? After all, they too are physical media transmitting 2D, coherence-enriched, recorded information. Like the Internet, they allow users to escape into alternate realities that are more structured than the real world due to the statistical requirements of legibility. As low-entropy representations, the stories transmitted by these media pose similar problems to human sense making. The power of narrative representations to impair one's ability to understand the real world and find satisfaction in everyday life has been explored by individuals as various as Socrates, Cervantes, Austen, and Joseph Gordon Levitt. By reducing natural variation to more quickly legible patterns of variation, narrative representations beguile our senses as we engage with them and possess the power to change how (or if) we notice natural variations and complexity moving forward. Once you have encountered enough fictional "bad guys," you might think they actually exist.

It is important, however, to emphasize that the problems posed by low-entropy representations (the form of recorded information that concerns us here) that we describe have more to do with the extent to which they substitute for live sources of information than with their low-entropy nature in and of itself. We predict that were novels and television to provide significant portions of an individual's informational diet, they could lead to leading indicator dependency or addiction. Moreover, we predict that such a state would correlate with degradation in a brain's model resolution and psychosocial effects similar to those we associate with Internet addiction. An extraordinary study from the 1970s in which researchers were able to measure how newly gained access to television changed the inhabitants of a remote Canadian town suggests that excessive TV watching impairs problem solving, perseverance, and the ability to tolerate unstructured time (Kubey & Csikszentmihalyi, 2002). It is estimated that 5% to 10% of the population suffers from television dependence (Sussman & Moran, 2013).

Novels, television, and the Internet are different, of course. Neither the novel nor television offers the same levels of interactivity, customization, or sense of place as the Internet, nor can novels or television conceal as deftly as the Internet the seams that reveal their fictionality. Advanced digital technology such as real time face capture and realistic digital avatars allows content creators to take to new levels the strategies of stylized, selective representation that make narrative representation possible. The Internet is the playground of an army of professional entropy reducers, all doing what they can for their own reasons to further simplify and compress the outputs of the others, creating ever newer mashups from ever more caricatured representations of lived experience.

As a physical medium, the power of the Internet to re-present reality in ways that beguile us inheres in its protean nature. Functionally, it is made of computers, fiber-optic and copper cables, and code. Physically, it appears to most people as a piece of glass in front of their faces. Whereas a codex or a television remains fairly fixed perceptually no matter the symbolic form (i.e., content) they display, the Internet (as opposed to a laptop screen, for instance) can change how it appears to users to suit a symbolic form. It may deliver the same symbolic forms one might encounter in a codex or a television set, say, a novel, film, or television program. Or it may deliver symbolic forms that were made especially for it—for example, social media feeds, blogs, video games, animated content visualizations, memes, podcasts, slide-share presentations, customer reviews. It can display text, still images, and moving images in various combinations. It can transmit sounds, both natural and artificial. In the future, it is likely to become even more protean. It solicits and responds to feedback in ways and at speeds that novels and televisions cannot, and does so with an agenda enforced by algorithmic intelligence. Our sensory systems are ill equipped to deal with a single "object" capable of responsively playing such disjointed roles. Novels and television offer attractive respites from reality. The Internet offers to replace reality with whatever we find desirable in a given moment.

What about stereoscopic virtual reality (VR) displays? Such displays allow a person to interact with a three-dimensional computer-generated representation (Cummings & Bailenson, 2016). The illusion of depth provided by VR would seem to be the solution to the problems we describe here. However, while VR can provide a sense of "presence"—"the subjective experience of being in one place or environment, even when one is physically situated in another" (Witmer & Singer, 1998)—this realism makes people sick. "Simulator sickness" or "cybersickness" as it is known, is thought to be due to "various stimuli, including linear oscillations at 0.2Hz, vection, visual distortion, flicker, conflict among oculomotor systems, and cue asynchrony" (Kennedy, Lane, Berbaum, & Lilienthal, 1993). Those problems were noticed decades ago and persist in spite of technological progress. It is unlikely that VR will become addictive, since users cannot tolerate it in the doses needed to trigger homeostatic fragility.

We do expect VR to become an important part of the 21st-century workplace, particularly as a tool for managing geographically dispersed teams, as with telepresence. Being 3D and immersive, VR has the potential to be richer than older communications technologies. Although our metrics are more refined than those used in media richness theory, we arrive at the same conclusion: richer media reduce uncertainty and equivocation more efficiently than less rich media. "Face-to-face is the richest medium," explain Richard Daft and Robert Lengel (1986), "because it provides immediate feedback so that interpretation can be checked." Daft and Lengel characterize richer media as using multiple channels and allowing for free expression and fast feedback. Our more refined metrics could be used not simply to match communications media to communicative task, but also to design VR with greater presence. Meetings held in a VR environment could benefit from the medium's ability to transmit nonverbal channels of communication, especially if those channels were modeled with spatiotemporal fidelity rather than as categories like "head and torso movements" (Hale

& Hamilton, 2016). It might be tempting to capture and algorithmically distill those cues as a visible bar or heat map, but such a process would suffer serious information losses during both encoding and decoding.

## 8  Research Directions

The two most radical aspects of this theoretical framework are its treatment of dysfunctional-internet use patterns as a normal, nearly inevitable reaction to statistically altered environments and its reliance on sound theoretical principles over uncertain data. Both aspects can aid in constructing a research framework of comparable reach. The key question to address in human-digital interaction is how we humans react when the structure of our social interaction is no longer arranged by our instincts but by the communicative needs of some intervening medium, and no longer arranged according to our interests, but according to its interests.

We begin with a few general suggestions, and end with testable hypotheses.

### 8.1  A Generalized Ecological Research Framework.

A framework for clinical research ought to provide a common, generalized platform with which different disciplines can communicate to study diverse effects in common terms. We recommend what is usually called an ecological approach, which examines how the environment affects perception, action, and cognition. Central to such an ecological approach in the case of dysfunctional use of interactive digital technology is the concept of informational diets. The informational diet concept could unify the multiple categories of digital use and abuse into a few simple parameters, and thereby promote the "cumulative function of research" needed for progress in Internet addiction (Tokunaga & Rains, 2016).

### 8.2  Human-Centered Theoretical Metrics

#### 8.2.1  Existing Metrics.

Media richness theory, as developed by Richard Daft and Robert Lengel (1986), explains how organizational structure can facilitate organizational information processing by ensuring effective communication between relevant parties and emphasizes the speed with which rich media reduce equivocation. Though the theory is still widely applied, it was developed before many of the communications technology commonly used in workplaces now. Moreover, the theory does not adequately take into account the ways in which media can be configured to increase or reduce richness (Dennis & Valacich, 1999). Our metrics could be used to refine and update this theory in terms of bandwidth, latency, and neuromechanical trust and may explain why workplace restrictions on e-mail can reduce stress and increase productivity (Mark, Voida, & Cardello, 2012).

*8.2.2 Theoretical Principles for Metrics.* Or course, data must be measured on limited schedules and budgets, putting a premium on cost and ease of measurement. But the historical success of physical and evolutionary law derives from concepts grounded in mathematical necessity, not in experimental ease. The solution is a common set of principles, perhaps masquerading as quantitative metrics, that appeal only to first principles and uncontested scientific fact, akin to the conservation laws of physics. This is the language not of experiment but of universal law. In culling concepts, even more crucial than excluding experimental categories is excluding social (as opposed to biological) measures of dysfunction. A truly human-centered framework must measure only health, unsullied by complications of legal punishment or economic reward.

As physicists have long boasted, universal laws are continuous and parameterized by very few dimensions (Feynman, 1965). The most successful big data projects exploit this fact by projecting high-dimensional discrete data, such as online click streams, onto low-dimensional curves (author W.S. has supervised several such commercial projects, whose details remain proprietary). In this view, collapsing the dizzying array of human-digital interactions onto a few key axes is in fact the only possible way to make sense of the disparate data documenting informational malnutrition.

While mathematically essential, such a truly continuous understanding of human informational needs renders unstable categorical concepts such as "dysfunction," "addiction," and even "diagnosis." We claim such categories do not in fact describe reality anyway, operating more as triage filters selecting where to spend resources. Researchers must take care not to confound categories imposed by convenience with measures of reality.

*8.2.3 Neuromechanical Trust as the Ur-Principle of Sensory Metrics.* We claim that the dominant mechanism of trust formation is equivalent to the mechanisms underlying the implicit trust in one's senses and one's balance. Such concepts are definable (if not already defined) in terms of stability zones and timescales and can be measured biophysically, so we need only point out a few consequences.

First, as proposed above, one can construct theoretically sound sensorimotor metrics, much like the metrics we propose valuing bandwidth, unfractured data, and precise timing. Second, one can rank the degree of trust a mind has in anything in direct proportion to the quality of its assumptions and the quantity of its accumulated data, so comparing trust in (say) a video image versus a live face ought to be straightforward. Third and most crucial, this principle when applied to human communication means that unconscious, nonverbal signals between people carry the vast majority of social meta-data underlying trust, and carry the bulk of emotional communication.

**8.3 Four Testable Hypotheses.** Our framework "predicts" many well-known facts: that walks in nature produce good moods, that people trust face-to-face conversation more than remote, that full-range motion is healthier than sitting sedentary. Such a continuous framework could of course divide its novel predictions in countless ways. We choose four general and testable hypotheses to show its scale and promise:

1. **Informational nourishment**. Human trust is maximized by distraction-free, long-duration interaction in natural multisensory 3D environments and degrades proportionally as those constraints are violated. We hypothesize that sensorimotor reintegration, if introduced incrementally, will provide the most effective recalibration for digitally impacted individuals.

2. **Resonant connection**. The best continuous approximation to human interaction is not bilateral communication but vibratory connection, as between coupled clocks or tuning forks. This generalized resonance would appear as a statistical correlation between multiscale, multidimensional, multifrequency attractor states (themselves composed of actively controlled superpositions of the classical mechanical eigenmodes of a solid body of minutely variable elasticity). We hypothesize that under ideal conditions, the microscopic interaction between two humans in proximity or contact will show traces of resonance and that interpersonal trust will correlate with its amplitude, duration, and salience For example, we predict that a mother's digital displacement during nursing will disrupt resonance and increase the baby's social anxiety.

3. **Proximity connection**. Under ideal conditions, the interactive trust between two people varies with their exchange of information, sensitive to both signal and noise. Among other things, this relation decreases with signal strength, and thus with distance, just like gravity or electrostatic force. The farther apart, the weaker the connection is. This hypothesis could be tested (say) by varying physical parameters (chair separation, lighting, air-conditioning noise, devices) in live meetings, or by comparing live interaction with teleconferences having various video and audio fidelities or latencies, for example.

4. **Neuromuscular healing**. All sensation, emotion, and cognition are built from neuromechanical interactions, and "stuck states" are built from microscopic neuromuscular spasms. We expect that local paralysis (e.g., Botox) therapies that dislodge facial spasms and emotions in fact work everywhere in the body. We predict the most effective therapies will involve training a body's most central, low-latency neuromuscular interactions: the proprioception and kinesthetics of spinal dexterity, balance, and vision, with particular emphasis on the midline core. We predict that after myofascial and kinesthetic training, mechanical improvements such as range (e.g., degrees of freedom, tracking precision, self-awareness) and grace (coherence, mechanical efficiency) will correlate with emotional and self-regulatory improvements.

## 9  General Discussion

We set out to answer two questions: Do informational diets matter? If so, how? We discovered that for deep mathematical reasons, live information streams are necessary to brain function.

Live information originates from 3-space and the people and things within it. Its interface is air. No matter their content, live sources are three-dimensional and constantly changing. Recorded information must be temporarily frozen for preservation or transmission. Such freezing violates the brain's statistical contract with nature. Furthermore, restoration of digitized signals necessarily mimics the appearance of continuity (as novels, television, and film can) by smoothing away compression errors, as our brains also do. No representation could ever preserve live information's raw density, and no transmission or interface could avoid interference. Interactive artificial representations further increase the sense of realism and presence, but thereby push quantization damage further into the unconscious, where it is harder to detect. For these and other reasons, we believe digital interactivity is incompatible with human sensory processing circuitry and damages social trust. This unorthodox claim could be refuted, for example, if experimental subjects (people or animals) emerge from multiday immersion in virtual reality undamaged.

Beyond the obvious inadequacies of digital sensorimotor data, we also find that their coherence- and discontinuity-enriched statistical structure can be potentially addicting to creatures evolved to process a more neutrally structured diet. Using foraging theory and control theory, we show how the natively stable, homeostatic feedback loop of a simple control circuit foraging for resources (informational or otherwise) could reverse its sign if the statistical distribution of the resources is enriched in attractiveness. In that case, as with addiction, pursuing short-term appetites exacerbates long-term needs.

This failure mode of leading-indicator dependency, when applied to selected, enriched information on the Internet, can drive our low-level informational appetites to consume ever more coherence-enriched stimuli in a doomed attempt to resolve the calibration problems posed by consuming too much in the first place. Such a decalibrated individual would display not only self-reinforcing consumption behavior but also well-known symptoms of Internet addiction or starvation (Baker & Kermidas, 2013) such as anxiety, irritability, depression, fatigue, and preoccupation with the stimulus. In short, informational malnutrition leads to informational starvation.

Our model of informational addictions is very similar to the impulse control model of Internet addiction, in which the Internet is viewed as a way to escape and felt as a source of instant gratification. Tolerance develops, as does withdrawal. The individual continues to overuse the Internet, despite awareness of long-term damage and despite efforts to reduce. Our model differs by locating the damage simply in the ratio of times spent online and offline.

We reject the category of "addict" as we reject all categorical representations. We prefer a continuous spectrum describing Internet decalibration. In its severe form, decalibration is a serious impulse control disorder whose prevalence rate is likely to increase as online "life" becomes more culturally acceptable. As with decalibration's intensity, its endurance must also lie on a spectrum, dependent on too many variables to count and in contradicting ways. Pressure at work, for example, might lead to online stress relief, or might remove the time for it. Damage to a relationship might provoke online recriminations, or provide impetus for interpersonal change. In general we predict that those who have ready access to supportive real-life interactions and diversely appealing multisensory environments, especially those motivated to change, will fare far better than those who do not.

To be sure, the Internet is not going away. It has certainly increased short-term productivity. Unfortunately, we have identified short-term feedback as problematic, and we have identified productivity as related to entropy reduction, which is the root of the problem. In economics as with software products, the largest efficiencies result from standardization, modularization, scale, branding, and optimization, all of which are mechanisms of entropy reduction. No wonder modern environments are so low entropy: natural variability has been straightened and smoothed into flat asphalt and clean desks, replaced by the unnaturally attractive variability of flashing billboards and flickering pixels.

To us, modernity's crucial distinction is that its most attention-getting features were designed to get attention, from skyscrapers to smartphones. Digital media, for example, have been economically optimized for short-term transmission, responsiveness, and attractiveness at the expense of fidelity. Digital media succeed by appealing to high-level recognition at the expense of the sub-millimeter, sub-millisecond detail that is so expensive to transmit. What is signal to our bodies is noise to a digital encoder. This matters because such detail carries the vast majority of sensory bandwidth necessary for interpersonal calibration and trust. Digital media succeed because they operate so fast and far, delivering long-range, short-term social hits but failing to deliver long-term trust.

Internet addiction is a complex, fast-evolving problem affecting practically everyone on earth; of course it is hard to study with experimental methods. Our model of Internet addiction, however, is both theoretical and highly simplified, features we regard as virtues. The sensory metrics we propose are commonplace yet yield orders-of-magnitude differences in information density. The sheer size of the difference renders precise numbers unnecessary, and its decade-scale acceleration suggests the Internet may be the straw that breaks the camel's back. Human environments have gradually enriched in attractive, low-entropy patterns since the first cave paintings and now, through the Internet, can dominate our sensory diets.

A natural instinct is to believe that humans can adapt to digital environments as we have to every natural environment on earth. But the Internet is neither natural nor on earth. It is not a source of informational nutrition but rather a source of informational lures and camouflage. There is no reason to think the human mind can adapt to a purely symbolic world as well as it can to forests and deserts or that younger generations benefit from childhood sensory fracturing.

Our hominid ancestors evolved millions of years ago to sense and communicate without technology or even words (which arrived only 50,000 years ago). Now even live words are ever more quickly being replaced by text communication thanks to a well-known dynamic called "network effects." The fact that social time is ultimately limited means that increasing time spent looking at screens decreases time spent looking at faces (see Figure 12).

Whatever their current flaws, digital media could improve in ways suggested by our metrics: reduce latency, jitter, and interruptions; increase bandwidth; and employ compression algorithms that do not privilege conscious recognition over unconscious feelings. We imagine a Web 3.0 that caters to rather than insults human sensory needs, organizations that encourage high-bandwidth interaction, and individuals who seek it for themselves.

This framework suggests that one key to healthy life with digital technology is systematic exposure to high-entropy natural stimuli. For example, the Japanese nature meditation of *Shinrinyoku* (aka forest bathing) has been shown to reduce stress, anxiety, depression, and anger (see the summary in Li, 2010; Tsunetsugu, Park, & Miyazaki, 2010). Other high-entropy treatments include proximity training (co-breathing, eye gazing), touch therapy with people or animals (Kaminski, Pellino, & Wish, 2002), and practices building neuromechanical trust (contact dance, partner sports). Following their own advice, we have invested in and benefited from related practices: autonomous motion (Ecstatic Dance), vibratory meditation ("Mindful Resonance"), and Mysore yoga.

We predict that neuromuscular interventions will succeed because the computational infrastructure of the human nervous system is more neuromuscular than cognitive. The most central, salient, high-bandwidth, and mission-critical signals in a body are internal: proprioception, kinesthesia, balance, posture, and locomotion. Next to those, even the visual and cognitive systems are but subsystems of head and gaze control. As a general principle, improving underlying muscular control produces cognitive and emotional benefits, as with eye-movement therapy (Shapiro & Laliotis, 2010).

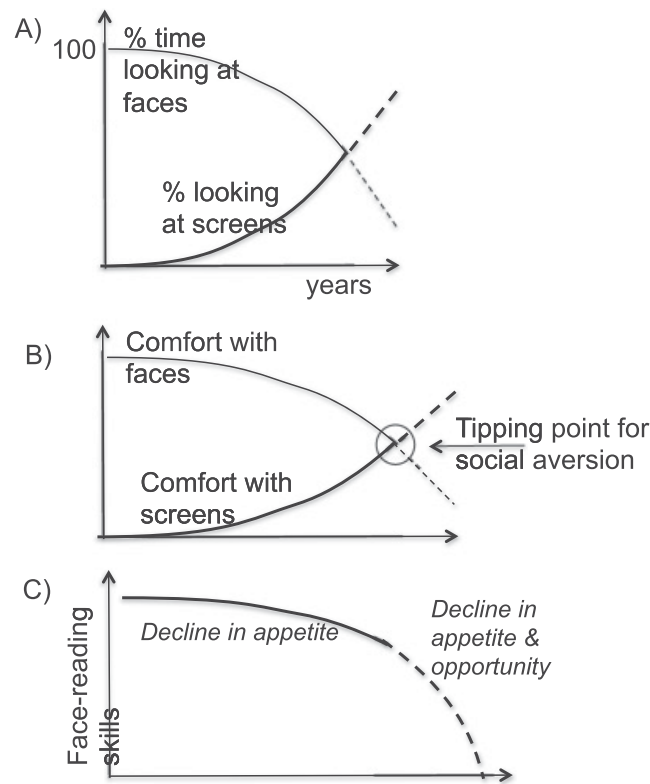

Figure 12: Growth dynamics of screen use. (A) The fact that social time is ultimately limited means that increasing time spent looking at screens decreases time spent looking at faces. (B) Because one's comfort with any medium increases (or decreases) with exposure and practice, there will be a tipping point at which screen use is preferred (about 6% of people worldwide now meet this standard). (C) While at first the decrease in face-reading skills is governed by individual appetites for screens, a sufficiently large screen-only population will deplete the face-reading population, leading to an additional decline in opportunity to see faces. These dynamics are mathematically neutral because they describe the general principles underlying progressive adoption of any communications medium, not just screens and faces.

Of all muscular systems, the most therapeutically promising is the most ancient one: the million-year-old spinal and breathing circuits underlying vertebrate identity. Massive anecdotal evidence indicates that spine-related practices such as dance, chiropractic care, and yoga routinely reverse lifelong muscular miswirings. Furthermore, the subtle pleasures of moving one's body through extremes of position and exertion are healthy because, as a general principle, all bodies evolved to meet their long-term needs through natural pleasures.

Like all other nervous systems, ours evolved to forage, not produce. Humankind uniquely produces things that captivate our senses, and now they do.

## Appendix: Framework Overview

This appendix explains and justifies our framework's concepts in quantitative terms. We occasionally repeat introductory remarks from the text for context.

### A.1 Continuous versus Categorical Representations.

Perception and learning in traditional neuroscience usually refer to categorical distinctions, such as object types, decisions, or behaviors—for example, differentiating friend from foe via assorted cues such as expression, vocalization, and gait. But a brain's twin tasks of millisecond muscular coordination and representing continuous reality require vastly more and better structured information than categories could ever provide. Here our computational, information-theoretic approach diverges from traditional neuroscience.

In our view, while categories can transport sensory information, they are bad at representing it because the natural world is continuous. Truly natural categories are few: the elementary particles and forces, atomic elements, and the states of matter. Typically, specific arrangements of matter follow statistical distributions instead; even friend versus foe is less a category than a heuristic to simplify ongoing, multidimensional responses to multidimensional inputs. Furthermore, quantized categories and decisions, shorn of raw detail, poorly approximate continuous 3D reality (a yes/no decision carries only a single bit, and one-of-many distinctions carry only a few bits).

Using continuous representation as a guiding principle makes some traditional measures unavailable. For example, human communication would be reduced to continuous bidirectional vibratory signals, more like sonic resonances than words. Words emerged only recently in evolutionary history. Doubtless among all that hominid humming and trembling, our nonverbal ancestors would have scattered proto-words and proto-gestures, but in bandwidth terms, that symbolic information content would have been only a filament within the continuous, semicoherent, multichannel carrier wave connecting two vibrating humans. In this continuous analysis, experimental measures like decisions, thresholds, and fixed-response questionnaires also have little place.

Even apparent natural discontinuities such as lightning and volcanic eruptions result from instabilities in continuous dynamical equations; so do electronic discontinuities like logic transitions, but far faster and more invisibly.

**A.2 The Unconscious.** It is beyond doubt that the vast portion of sensory and sensorimotor processing is inaccessible to conscious recollection, for two main reasons. The first computational limit on consciousness is horrific undersampling: sensory neurons feed the brain information a million times faster than it can store it. (For an accessible overview of the research on unconscious processing, see Mlodinow, 2015.) More crucial is that sensory awareness itself must operate unconsciously to be effective. Neuroscientists for years have marveled at the deftness by which the visual system hides the retina's blind spot from awareness, but they seldom note that as a general principle, such cloaking is a computational necessity for a representation to operate as quickly and efficiently as possible (Softky, 2014). Consciousness, like self-auditing, is usually a waste of computational resources.

Consciousness (which we do not model) is related to attention (which we do). The part of attention we are conscious of is the goal-directed attentional system, but there is a stimulus-driven attentional system as well (Eysenck, Derakshan, Santos, & Calvo, 2007), which operates below the conscious threshold. When the unconscious stimulus-driven system dominates, conscious attentional control decreases. We show that this unconscious attentional appetite is biased toward consuming low-noise "clean" stimuli, and we show the conditions under which too many clean stimuli can trigger avoidance of high-noise stimuli leading to compulsive Internet use. Unfortunately, digital media have evolved to please the conscious mind, so by construction we will not sense the unconscious damage they produce.

**A.3 Compression and Decompression Occur in Stages.** Those who frequently use digital devices are already familiar with the central concepts underlying compression, in particular encoding, transmission, and decoding of streamed images, as shown in Figure 2.

Take a familiar video camera, for example, and its several information-processing stages. First, the lens focuses a continuous two-dimensional sheet of light on sensors of the camera's focal plane, where in the process called encoding or compressing, light intensity becomes recorded as fixed, quantized values at fixed, quantized locations. Those bits in one form or another are stored and transmitted to a screen, where decoding or decompressing circuits interpolate them to conjure approximations of the smooth surfaces, straight edges, and continuous motion which were present in the original sheet of light. That reconstituted model drives yet another quantization step, flashing discrete pixels on a screen. When our eyes see those pixels, our visual system performs its own decompression, conjuring in our minds a sharp, glitch-free impression of the original scene based on that smattering of flickering dots.

All of this can be understood as the flow of Shannon information along a bandwidth channel connecting encoder and decoder, a process clearly more complex than mere bandwidth. It has several stages, as shown in

Figure 2. An encoding circuit must validate the format of its input data (stage 1) and set the meta-data about encoding context (stage 2), filter (stage 4), quantize (stage 5), and then, during decoding, validate the received quanta (stage 6) and adjust dequantization filters based on separately decoded context meta-data (stage 7), before inverse filtering (stage 8) and dequantizing (stage 9).

A signal can be called deceptive if it has been specifically encoded (in violation of that assumption) to exploit an adversary's hard-wired decoding assumptions of neutrality. For example, the high-contrast visual contours found in camouflage, or lures, exploit a viewer's automatic, low-level assumption that high-contrast contours correspond to object borders. In this scheme the middle stages, 3 to 7, correspond to information compression as it is understood for images, videos, and so on, while the outermost steps deal with meta-data. The middle data encoding and decoding stages are in fact the only ones implemented in systems optimized for efficiency; self-auditing and incorporating meta-data such as context slow things down and are thus omitted. For example, in the natural world a visual system would be optimized to process sensorimotor data directly, assuming for convenience and speed that all of its sensorimotor input in fact originated from 4D space-time. In a natural sensory channel, signal trust is necessary, fast, implicit, and preconscious.

**A.4 Bandwidth.** The quality of a representation depends on its information content or rate in bytes per second, so sensory bandwidth limits the quality of sensory representations. The lowest bandwidth senses are smell and taste because they change so slowly; we leave them out of our analysis. In descending bandwidth order, our top three channels are touch, vision, and sound. Estimates of bandwidth for those senses depend on whether the neural spikes are treated separately in a pulse code, or averaged together in a rate code, as shown in Table 1.

**A.5 Noise.** Variability in nature is noisy and random and can be averaged into better signals. Variability in the digital world was sculpted just for us; the structure of its variability either takes advantage of or prevents our signal-averaging abilities.

Before discussing signals, we should discuss noise. Noise in nature is abundant, and simple. At the very finest scale, photons and phonons are independent and random (Poisson shot noise). This means that the central limit theorem (aka law of averages) applies, so averaging provides a good estimate of the original source. Furthermore, statistically significant deviations from uniformity correspond to real deviations in source structure. This is the usual case for scientific data gathering.

But variability in the digital world has a very different structure from the "noise" known to science. In one sense, digital variability is lower, having been specifically enriched to appear to our sensory systems as coherent 3D images or sounds rather than as random snow or hiss. In that sense, moment-to-moment digital inputs are designed to seem low noise and clean. But digital sources are hyperdimensional patterns, which (unlike real things) can change discontinuously, thereby violating the continuous natural laws a nervous system expects. The unnatural structure of digital variability can make it appear far more trustworthy and predictable than it actually is.

Table 1: Approximate Bandwidths of Human Sensory Input.

| Source | Fibers | Spikes/ Sec/Fiber | Bits/Spike | BW (bits/sec) |
|---|---|---|---|---|
| Optic nerve | | | | |
|   Pulse code | 1,000,000 | 1 | 10 @ 1 msec pulse | 10 Mbit |
|   Rate code | | 1 | $0.15$ @ $N = 10$ rate | 150 kbit |
| Myo-fascial network | | | | |
|   Pulse | 1,000,000 | 1 | 13 @ 0.1 msec pulse | 13 Mbit |
|   Rate | | | $0.15$ @ $N = 10$ rate | 45 kbit |
| Cochlear nerve | | | | |
|   Pulse | 30,000 | 10 | 13 @ 0.1 msec pulse | 4 Mbit |
|   Rate | | 10 | $0.15$ @ $N = 10$ rate | 45 kbit |
|   Nyquist | | | | 10 kbit @ 20 kHz |

"Noise" in neural signals is problematic as well. Neural signals are composed of pulses (action potentials), whose timing seems roughly random ("Poisson"). As shown in Table 1, one can either average those pulses (the usual neuroscientific "rate code") or treat them as precise binary messages ("pulse code") (Softky, 1995). The difference amounts to a factor of 100 in information content.

**A.6 Signal Formals and Metrics.** Table 2 quantifies the sensory metrics contained in Figure 3.

**A.7 Media and Interface Distortions.** An interface lies between the organic sensorimotor system and the calibration field. Originally that interface was air. Air is neutral and consistent, transmitting signals with maximum fidelity along each of the eight metrics of Figure 3. Relative to it, digital communications are corrupted in many ways—in particular, decreased bandwidth, lower temporal fidelity, and more interruptions.

One can understand digital communications technologies like those of Figures 1 and 2 in terms of those interface metrics and can thus quantify artificial social communication alongside natural communication. Figure 13 shows how four digital metrics crucial to social communication—bandwidth, latency jitter, fracturing, and uninterrupted duration—all fare orders of magnitude worse than natural ones.

Table 2: Sensory Metrics of Trust.

|   | Metric | Units | Worst | Best | Importance |
|---|--------|-------|-------|------|------------|
| 1 | Encoding format | Yes/no | No | Yes | Wrong format violates protocol, allows deception |
| 2 | Unfractured sensory pairings | [1/25,1.0] | 1/25 | 1.0 | Cross-sensory inconsistencies violate protocol, mislearn correlations |
| 3 | Round-trip latency | Sec | Days | Microseconds | Fast round trip allows quick iteration and exploration |
| 4 | Latency jitter | Sec | Days | Microseconds | Consistently precise timing allows high-precision correlations |
| 5 | Uninterrupted duration | Sec | Sec | Years | Long calibration runs work best |
| 6 | Sensory bandwidth | Bits/sec | Bits/hour | MB/sec | Learning rate and performance are proportional to bandwidth |
| 7 | Noise | Fraction of signal | ≫signal | 0 | Noise reduces the number and bandwidth of detectable signals |
| 8 | Unbiased input fraction | 0–100% | 0% | 100% | Systematically biased input at all threatens calibration dynamics and thus security |

One major source of media distortion is the gap between apparent and actual information transfer. The apparent information is given by the resolution of the transducer (a display or speaker), which physicalizes uncompressed information. The compressed information feeding that representation is typically much smaller, as shown in Table 3.

*A.7.1 Example: Latency Jitter Reduces Trust.* Consider for the moment the brain as a sensitive correlation detector, able to correlate output spikes with subsequent input spikes over long times yet with millisecond precision and reliability resulting from the physical constancy of air's sound-speed, of the eyeball's moment of inertia, of retinal cells' response lag, of myofascial tissue's vibrational phase velocity, and so on. Such a brain would be sensitive to millisecond-level discrepancies in return signals from any source. So once a brain sends a message to an object, it keeps track of exactly when the return signal (echo) arrives and ascribes any variability not to the transmission medium but to uncertainty about the object itself, whether that "object" is another person's attention or one's own leg. This is why the metric of latency jitter is a good marker for neuromechanical trust, and hence for trust in general.

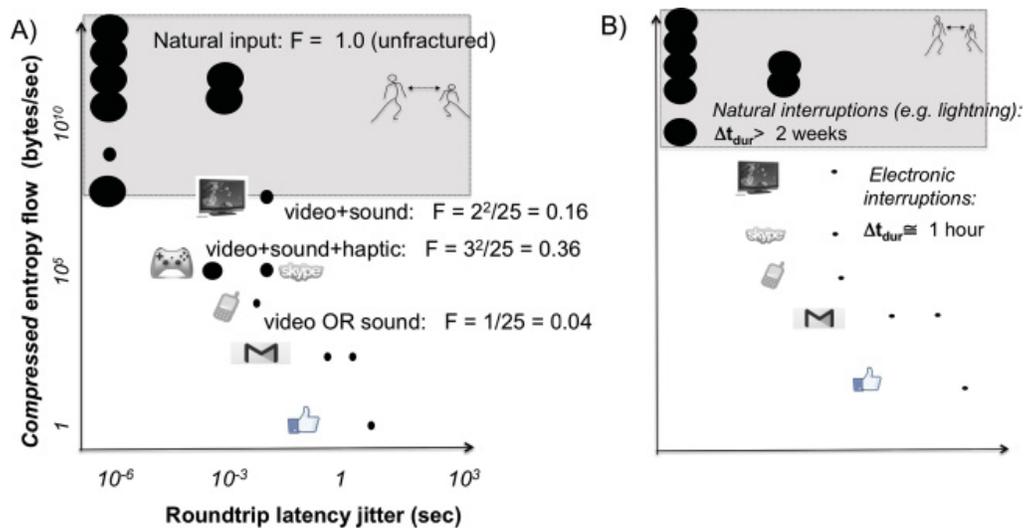

Figure 13: Quality metrics of natural versus digital stimuli. Entropy flow (bandwidth) goes down, and latency jitter (a measure of temporal uncertainty) gets worse with digital communication. Uninterrupted duration also gets shorter (same scale). Both effects span orders of magnitude. These graphs use the same layout as Figure 14 in plotting a scalar dot-size according to bandwidth and latency jitter, with the largest dots representing natural values. (A) Fracturing F measures the number of cross-sensory pairs possible in the medium, relative to the 25 pairs possible with five senses. (B) Uninterrupted duration $\Delta t_{dur}$ measures the upper bound of time available without interruptions. Digital media fare far worse than unimpeded access to the natural calibration field on all metrics, not just these two.

Table 3: Compressed and Uncompressed Bandwidths of Digital Devices.

| Message | Compressed Information or Bandwidth (bytes) | Uncompressed Transducer Information or Bandwidth |
|---|---|---|
| Facebook "like" | 1/8 (=1 bit) | 1000 (= 1000 pixels) |
| 1 character | 1 | 1000 |
| 1 emoticon | 3 | 1000 |
| 5-character string | 5 | 5000 |
| 5-character string (autocorrected) | 2 | 5000 |
| 162 character text | 162 | 16200 |
| 200 word e-mail | 1000 | 200 kB |
| Poor mobile call | 1 kb/sec | 5 kb/sec |
| Good landline call | 5 kb/sec | 5 kb/sec |
| Skype | 100 kb/sec | 10 MB/sec |
| HDTV | 1 MB/sec | 1 GB/sec |
| Telepresence | 4 MB/sec | 1 GB/sec |

Now consider what this means for digital communication. We suppose two humans communicating in proximity exchange reciprocal timing signals with precise (i.e., millisecond) phase precision. And we suppose their brains not only recognize but learn to rely on that high level of temporal response fidelity, albeit unconsciously, as a marker of the reliability of the other person, and hence as a marker of trust. By this measure, an exchange of text messages lasting an hour on average (3.6 million msec) has made each participant seem to the other to be millions-fold less trustworthy than in real life, at least unconsciously. Even a better and more highly interactive technology, the mobile phone, does not actually transmit millisecond-level signals but instead synthesizes them and, additionally, introduces a hundred-fold more jitter, becoming a hundred times less trustworthy than real life. In both cases, the drop in trust is assigned instinctively not to the medium but to the person at the other end.

**A.8 Natural Statistics.** The neuroscientific concept of natural statistics distinguishes the fractal complexity of natural scenes from the simplified structure of man-made cages and buildings; we explore this distinction in depth in section 5.

Most stimuli in visual neuroscience were designed to elicit spikes, for example, high-contrast shapes and gratings suddenly flashed on a screen or symbols carrying fixed semantic content. Those stimuli produce such strong neural responses precisely because they are unnatural.

By definition, the statistical profile of the natural world is "natural." The term *natural statistics* conventionally refers to the statistics of natural visual features, and we follow that tradition in assuming that the Bayesian priors embedded in sensory systems are hard-wired for the statistics of natural environments. But we disagree about measurement, which for experimental convenience has analyzed images of nature, that is, 2D freeze-framed pixelated projections (Ruderman & Bialek, 1994), rather than the ever-changing continuous four-dimensional original sources, such as waving trees or swirls of smoke. The rich, multiscale detail of natural 4D environments makes them distinct from hypersimplified environments such as the rectilinear cages in which lab animals acquire most of their sensorimotor experience. While both natural and cage environments deliver continuous-time 3D signals, their statistical profiles differ profoundly, with cage environments missing crucial high-entropy signals. In effect, the same reasoning that makes nervous systems reared in cages suspect should also apply to nervous systems reared in buildings.

*A.8.1 Quantitative Comparison of Natural and Artificial Inputs.* Comparing bandwidth values across digital and natural environments requires a thought experiment in which the same stimulus can be expressed in both terms. We introduce the concept of a flatland video display: like a screen, but a continuous-space, continuous-time, full-color, full brightness, full-resolution image, like the real world in all respects but flat instead of 3D. We make the equivalence by equating the information flow necessary to drive such a display with the subset of sensory experience that perceives it. Figure 14 illustrates the comparison.

Specifically, in a flatland display, a viewer sees not a monocular image of a 3D person, but a binocular image of a 2D projection of the person. That image would span only a small portion of a viewer's visual field, and none of her tactile or kinesthetic field. Furthermore, by being flat, it would carry even less information. So in natural terms, such an image is information poor. By circumventing natural depth cues from eye-vergence, binocular disparity, and interaural timing difference, this flatland image would not only fracture sensory experience; it would occupy only a small fraction of the viewer's 3D representational space and thus consume only a small fraction of its native 3D sensory bandwidth.

While such flat images are suboptimal images for our senses, in technological terms they constitute an ideal, infinite-resolution telepresence display. So this hypothetical sub-lower-bound example of a seminatural calibration input would already exceed the upper bound of any imaginable technology, making the bandwidth comparison approximate at best. But because we want to compare natural and artificial communications channels, we can choose to equate the sensory bandwidth of "flatland" to the known telepresence bandwidth of about 1 gigabyte per second. In this sense, even telepresence counts as sensory deprivation. It is important to emphasize that text-based channels carry yet a billion-fold less information than telepresence.

**A.9  Calibration as a Process.** Instruments are calibrated by engineers and scientists to maintain resolution, noise immunity, and robustness. But strictly speaking, calibration applies less to the physical instrument than to a model of its input/output relations. So in the broadest sense, the goal of calibration is to produce a trustworthy model, operating over as wide a range and with as small an error as possible. In this sense, a trustworthy model would thus roughly maximize model trust, which we represented relationally in Figure 4.

In other words, calibration activity is a form of data gathering designed to make the instrument's behavior match its current resolution. Both follow a process beginning with the initial creation of high-amplitude excursions over a large volume of data space and followed by the slow reduction in amplitude and speed as the model's resolution improves. The better the model, the more subtle the signals it seeks, the longer it consumes them, and the gentler its output perturbations.

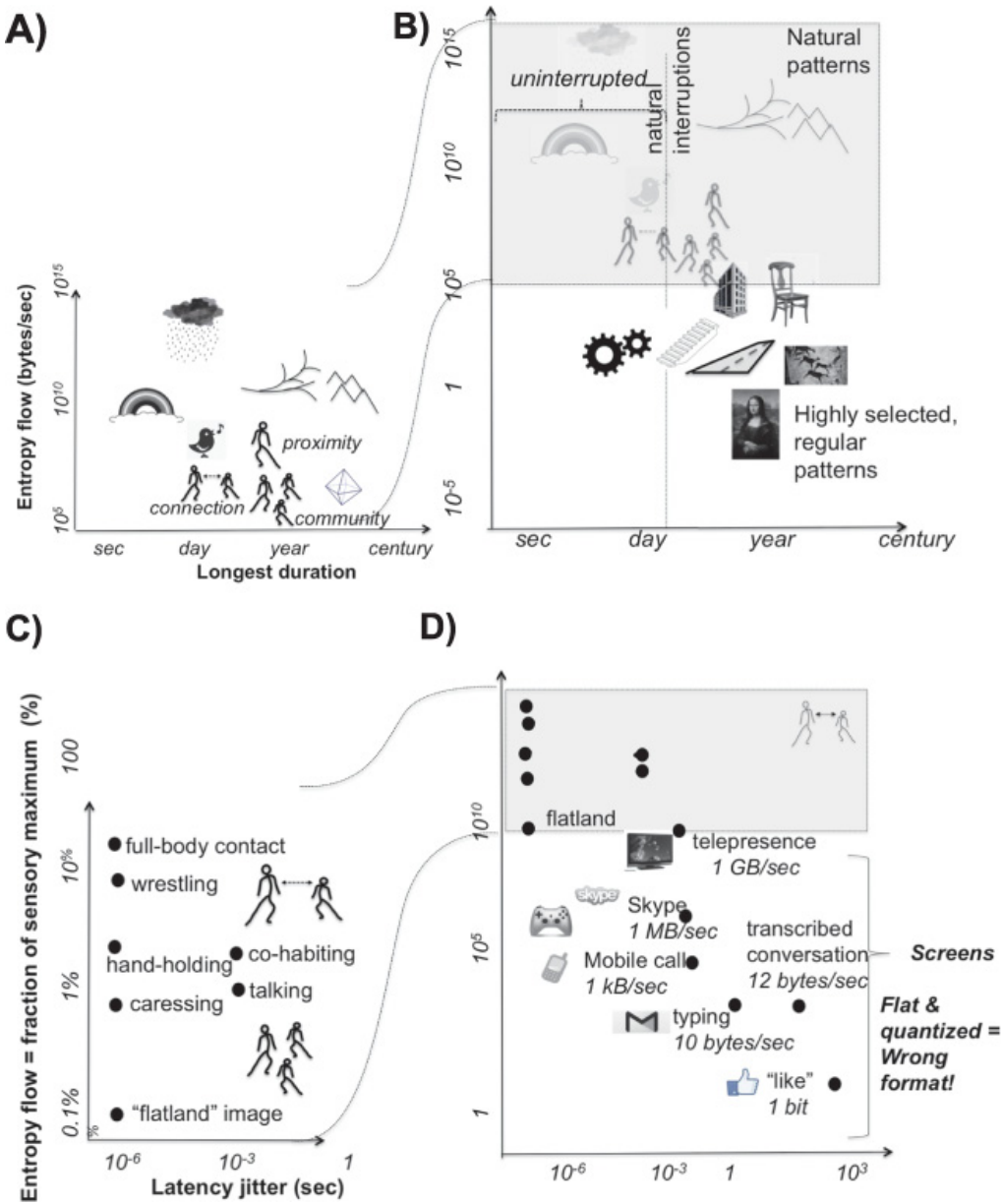

This process is so familiar that it is seldom discussed, but the rough steps and rationales of this approach bear repeating because they tightly constrain how self-calibration must proceed. The qualities of good calibration data are familiar, but each calibration stage requires a different type of data. We represent the relationship between calibration stage and data type in the following list in order of decreasing output amplitude:

1. **Orient.** In an interactive system like a sensorimotor system, sensory inputs must respond to outputs before testing can begin. The system must first orient itself by creating coherent, commensurate output and input filters, as with rocking a radio dial to optimize a signal (a process often called "locking in").
2. **Probe extremes of input and output.** Before beginning data storage, one must know at least the data's approximate range, and hence must push signal amplitudes at least briefly to their limits.
3. **Accumulate statistics of the interaction.** Once a stable interaction is established, the best calibration runs are the longest uninterrupted ones with the widest output ranges, including no output at all (neutral baseline). Not only are more data better as a general principle, but uninterrupted data in particular avoid introducing uncertainties regarding context switching, reestablishing lock, and reverifying parameters.

Initially, to lock in a signal quickly and test extremes, it helps if the signal is underambiguous (i.e., "clean") in some domain, in the manner of a test

---

Figure 14: Natural patterns sorted by entropy and duration. (A) Spatial patterns in the forest or savanna are ambiguous, complex, detailed, and diverse—of high entropy. In that environment simple low-entropy patterns are rare and often fleeting, but those provide the cleanest signals. (B) Man-made patterns have been selected over time and, hence, contain more regularities than natural ones. (C, D) Channels of nonverbal (continuous) and symbolic (quantized) communication: (C) Mechanical vibrations of the body carry the highest-bandwidth interpersonal communication signals, involving up to 100% of the sensorimotor system (as in full body contact) down to very little (grazing touch). The horizontal axis here shows one possible measure of sensory information quality, the jittering in latency (equivalent to the signal's temporal precision). In a natural environment such latencies are effectively instant (except for sound speed) and jitter free. As a thought experiment to compare natural with artificial inputs, one can image a two-dimensional flatland image embedded in three dimensions, which would occupy a tiny fraction of volumetric sensory input and thus would carry far less bandwidth. (D) The same bandwidth analysis applies to standardized communications like videos, phone calls, and text, with the crucial proviso that none of those actually arrives in the continuous space-time format a nervous system expects. The uncompressed bandwidth of those media depends on the transducer (e.g., screen) resolution, but their compressed (i.e., true) bandwidth is consistent. To present vibratory and quantized communication on the same axis, we equate the flatland image from above with an ultra-high-quality telepresence interaction. This approach shows the impoverished form and quality (via lower entropy and higher latency jitter) of electronic communications relative to natural ones.

pattern: sharp contours, sudden transients, pure tones, or pure hues. Generalized edges and point sources are especially useful because they carry information phase-locked across multiple scales. Successful signal locking does not last long, but subsequent steps do, because more data always give better statistics. Slowly accumulated signals should be natural (as opposed to coherence enriched) and contain a representative amount of ambiguity and noise. In summary, the general process of calibration usually moves from brief, coherent, high-amplitude outputs and input to slow and subtle ones, because calibration usually improves the resolution of the model. But if the resolution somehow gets worse, proper calibration will reverse that strategy, boosting output to get clean return signals.

**A.10  Palatability.**  We assume informational organisms will seek to recalibrate by seeking clean, unambiguous signals in natural environments. We map this appetite for clean sensory data to the concept of attractiveness or "palatability."

Artificial digital environments are enriched in overcoherent signals to meet informational appetites. But much like lures and living fish, artificially attractive signals cannot satisfy an organism's underlying needs. Digital sensory data in particular fail to meet human needs because they do not originate directly from the 4D space-time for which we evolved, but arrive in an utterly different data format: they are two-dimensional, discontinuous in space and time, and delivered at low bit rates (relative to physical proximity). A human consuming such quantized input, no matter how clean, will remain hungry for the rich, continuous, 4D signals required by human sensory processing circuitry. When the informational need for continuous 4D signals remains unsatisfied, processing circuitry falls out of calibration, and the accuracy level of the brain's model of its world declines. In a doomed effort to restore calibration, sensory processing circuitry may drive a human to seek the easily accessible yet unserviceable clean data provided by digital environments.

**A.11  Sign-Reversing Feedback Loops.**  When organisms operate outside the statistical environments in which they evolved, their expected homeostatic balance can be upset, and stable negative feedback can be inverted into positive feedback in a situation called "sign reversal."

Since the evolution of dopaminergic reward signals, organic circuits have evolved to control various biological parameters such as nutrient intake, energy expenditure, salt concentration, and so on. These circuits are what keep organisms in homeostasis as their environments change. Two conditions allow such circuits to function even in the presence of sensor and effector flaws, such as time lags, noise, and imperfect correlation with the ground-truth biological need. First, homeostatic control is by construction self-correcting, so it should converge even with flaws in the sensors and effectors. Second, such control succeeds by following concentration gradients through space, so trial-and-error tuning (i.e., evolution) is sufficient to optimize the circuit parameters for any particular combination of sensor properties and concentration statistics.

But as a general rule, circuits optimized by trial and error to operate in one type of environment will fail, sometimes catastrophically, in environmental structures they have not experienced. We examine concentration gradients because they are the organism's eyes; they show it where to go and what to do, and their structure is embedded in the organism's circuits and parameters. If modern gradients point differently than evolutionary ones do, not only does the original homeostatic contract no longer hold, but now a highly sensitive automated vehicle is on the loose unsupervised, allowed (or forced) to both forage and self-calibrate in completely uncharted zones of data-space, using only old hard-wired habits but otherwise without references. It is hard to conceive of any circuit that would not get stuck or fail under such circumstances.

**A.12 Evolutionary Pressures.** The most general problem is that humans evolved to sense and communicate without artificial interference. Only very recently in evolutionary time, at the "great leap forward" about 50,000 to 100,000 years ago, did physical artifacts such as tools and cave paintings begin filling an ever-growing portion of the human calibration field (Diamond, 1999). Unfortunately, the laws by which physical messages propagate depend not only on human salience, but also on physical productivity, so over time, unproductive or antiproductive messages die out, biasing the message pool away from human-centered distributions. This line of reasoning poses several systematic tensions, and explains Internet addiction as simply the latest and most potent calibration imbalance caused by various representational technologies over history.

The first tension concerns conscious versus unconscious recognition: digital inputs are optimized for conscious recognition. Yet we cannot detect (much less correct for) problems at scales too fine or fleeting to notice. The second tension concerns discomfort versus dynamic range: many of the torments that civilized life avoids—thermal stress, hunger, violent illness—can be considered in informational terms as unpleasant but necessary calibration episodes. Third, the attention-harvesting industry exploits the tension between short-term compulsion and the long-term mental externalities it creates. The fourth tension concerns network effects: the more people abandon our native face-to-face protocol in favor of convenient digital interactivity, the fewer live bodies remain for others to practice with (see Figure 12). If left unchecked, this trend will ultimately make obsolete homo sapiens' native communications protocol.

Perhaps the deepest systematic tension is the tug of war between optimization and autonomy, both of which are needed for establishing any kind of trust. Like life itself, optimization is a form of entropy reduction. Evolution optimizes creatures by natural selection; evolution optimizes brains to perform efficient representation; a growing brain optimizes its own performance through learning; and a mature brain optimizes its balance and body coordination using data in real time. Paradoxically, each entropy-reducing optimization includes a step of injecting entropy, like autonomous choices and random mutations.

Businesses also optimize: a business as a whole optimizes marginal revenue, data science finds best-fit models, and websites iteratively improve click counts by random A/B testing. Unfortunately, optimizing a business goal often involves sculpting customer perceptions and behaviors, thereby reducing the entropy of human behavior, choice, and autonomy. In short, to calibrate our brains, we need autonomy; to submit it to organized optimization, we must forgo it.

## Acknowledgments

We thank our anonymous reviewers for their incisive feedback on our original submission; in particular, we are grateful for the prompts to provide more detail in the control theory section and to add a section on physical media and alternate realities. We are also grateful for helpful comments in person and print from Adam Safron, Eric Doehne, John Joss, Camille Everhart, Kwabena Boahen, Tatiana Engel, Bill Newsome, Tyler Holcomb, and Alex Anderson, and for discussion with Boahen's Brains in Silicon group and Berkeley's Redwood Center. Both of us conceived of and developed the ideas presented in this note drawing from our respective expertise and training.

---

**Author's Post-Publication Note:**

The authors intend to release a clarification of the block diagrams in section 6.

**FOR IMMEDIATE RELEASE**

**A Neuroscientist and a Humanist Find Simple Solutions**
**to "Internet Addiction" Using Math**

**MENLO PARK, CA (August 31, 2017)** — Research in both the natural and social sciences finds that the more time people spend online the more likely they are to be lonely, depressed, and anxious. The nature of this relationship, however, remains poorly understood. To solve this mystery, theoretical neuroscientist William Softky and narrative theorist Criscillia Benford turn to the mathematical laws of information flow. Their paper, "Sensory Metrics of Neuromechanical Trust," which appears in the September 2017 issue of the *Journal of Neural Computation,* uses these laws to model how sensorimotor information flows through the circuitry responsible for calibrating the brain's internal representations of lived experience. This circuit-level model demonstrates how recorded information (like digital data) negatively impacts the brain's powers of prediction, causing humans to lose trust in their senses, themselves, their understanding of the world, and each other.

Softky and Benford treat all incoming information—whether information from a wilting rose or information from the physical presence of a live, talking person or information from a text exchange or Instagram photo, etc. etc.—as training data for the brain. Training data originating from digital worlds negatively impacts human understanding and trust because it violates the brain's hard-wired expectations regarding signal format, noise, and timing; it elevates vision at the expense of other senses; it contains less information (thanks to dominant compression practices and physical limits on bandwidth) than data originating from the "real" world; and it is systemically biased in ways that generate revenue. When posed in these terms, the problem of "internet addiction"—or as Softky and Benford prefer call it "leading indicator dependency"—is best solved through the systematic reintroduction of live training data originating from "woolly" physical environments like forests, and woolly social situations like face-to-face conversations on sensitive topics.

Live training data from woolly places and situations might feel uncomfortable to process in the short term, but in the long term it makes people feel better, because such data is trust-bearing and ambiguous enough in just the right ways to re-calibrate a brain's internal models. As Softky explains, "The human brain understands the trustworthiness of its models like a scientist understands the trustworthiness of calculations. Both brains and scientists assign 'confidence intervals' to every data point, which are shadow or mirror numbers that say how much to trust that data. For example, the brain assigns a high confidence interval to signals from deep inside our muscles. This is why we experience neuromechanical trust when our foot hits the ground. Neuromechanical trust isn't inspired by 'content,' it arises from a separate channel carrying meta-data, a 'trust channel.' Our brains create trust from raw trust-bearing signals, which of themselves are high-bandwidth, unbiased, coherent sensory information from natural 3-D sources, with fast back-and-forth interactivity. Unlike recognizable 'content,' trust-bearing signals are processed below the threshold of consciousness. We don't always notice when trust signals are missing, but the damage is very real."

The paper addresses the most common question asked of people who investigate so-called internet addiction: Won't humans learn to handle the effects of digital technology on the brain? After all, as Benford points out, novels, TV, early video games and more inspired similar moral panics. "Socrates worried that the written word would make people more forgetful, stupid, and boring. It's easy to laugh. People adapt to the latest technology and the world wags on, right? That's the standard refrain," she says. "But consider this: our newest representational technologies are ubiquitous and necessary. They 'think' a million times faster than us, in microseconds instead of seconds, and they can relentlessly pursue our attention and our money with algorithmic intelligence. We've never been here before. Representational technologies have never before known so much about us, been so smartly integrated into daily life, so interactive, and so focused. The point is not that digital technology is de-facto evil. The point is that it can be—and must become—better tuned to our own needs as social beings with sensorimotor needs."

Benford and Softky hope the framework they've introduced will inspire "Web 3.0"—a wiser version of the internet that respects human sensorimotor needs—and facilitate the development of effective and accessible treatments for the psychosocial distress associated with problematic internet use.

Their mathematical model of what happens to the brain when it consumes high quantities of training data originating from interactive digital technology yields clear advice for individuals who are looking to cut down on their internet use and avoid communications minefields:

- Practice **tolerating uncertainty** by waiting to receive a definitive answer to your questions. When your physical safety is not at stake, give yourself the chance to feel the discomfort of being unsure. If you combine this practice with the practices described below, cravings for instant certainty will probably feel less intense.

- Practice **sensory integration** by making time for whole-body sensorimotor experiences, ideally in "woolly" environments: hiking, yoga, swimming, dance, forest bathing, gardening. Autonomous activities, especially outside, provide the full-spectrum training data that fuels trust in your senses. Solo sensorimotor practice is especially useful for people overwhelmed by face-to-face interaction, because it facilitates sensorimotor re-calibration without requiring social exposure.

- Practice **proximity communication without technological interference**, ideally for an hour at a time, in "woolly" environments—e.g., community picnics, family dinners, dances, sports, games. Close-up interaction provides the full-spectrum sensorimotor training data one human needs to trust another.

- **Choose the right communications medium**. Different media have different pros and cons. Email and text are excellent for transmitting concrete facts, but problematic for transmitting complex ideas, and horrible for transmitting emotions because they leave out cues our nervous systems need. High-bandwidth phone conversations are better for understanding and empathy, telepresence even better, with face-to-face the best of all.

**CONTACT INFORMATION:**

**Corresponding Author:**    William Softky
                             Bioengineering Department, Stanford University
                             wsoftky@stanford.edu